\documentstyle[12pt,bezier]{article}
  \input FEYNMAN
  \newlength{\extralineskip}
\newcounter{equnum}[section]
\def\theequnum{{\rm\thesection}.\arabic{equnum}}
\newcommand{\beq}{$$ \refstepcounter{equnum}}
\newcommand{\eeq}{\eqno (\theequnum) $$}
\newcommand{\bd}{\begin{displaymath}}
\newcommand{\ed}{\end{displaymath}}

\def\e{\, {\rm e}}
\def\IR{{\rm I}\!{\rm R}}
\def\inbar{\,\vrule height1.5ex width.4pt depth0pt}
\def\IC{\relax\hbox{$\inbar\kern-.3em{\rm C}$}}
\def\IR{\relax{\rm I\kern-.18em R}}
\def\IZ{\relax\ifmmode\mathchoice
{\hbox{\kern-.4em Z}}{\hbox{Z\kern-.4em Z}}
{\lower.9pt\hbox{Z\kern-.4em Z}}
{\lower1.2pt\hbox{Z\kern-.4em Z}}\else{Z\kern-.4em Z}\fi}

\def\tr{{\rm tr}}
\def\ps2{{\bar{\psi}\psi}}
\makeatletter
\newdimen\normalarrayskip              % skip between lines
\newdimen\minarrayskip                 % minimal skip between lines
\normalarrayskip\baselineskip
\minarrayskip\jot
\newif\ifold             \oldtrue            \def\new{\oldfalse}
\def\arraymode{\ifold\relax\else\displaystyle\fi} % mode of array entries
\def\@arrayskip{\ifold\baselineskip\z@\lineskip\z@
     \else
     \baselineskip\minarrayskip\lineskip2\minarrayskip\fi}
\def\@arrayclassz{\ifcase \@lastchclass \@acolampacol \or
\@ampacol \or \or \or \@addamp \or
   \@acolampacol \or \@firstampfalse \@acol \fi
\edef\@preamble{\@preamble
  \ifcase \@chnum
     \hfil$\relax\arraymode\@sharp$\hfil
     \or $\relax\arraymode\@sharp$\hfil
     \or \hfil$\relax\arraymode\@sharp$\fi}}
\def\@array[#1]#2{\setbox\@arstrutbox=\hbox{\vrule
     height\arraystretch \ht\strutbox
     depth\arraystretch \dp\strutbox
     width\z@}\@mkpream{#2}\edef\@preamble{\halign \noexpand\@halignto
\bgroup \tabskip\z@ \@arstrut \@preamble \tabskip\z@ \cr}%
\let\@startpbox\@@startpbox \let\@endpbox\@@endpbox
  \if #1t\vtop \else \if#1b\vbox \else \vcenter \fi\fi
  \bgroup \let\par\relax
  \let\@sharp##\let\protect\relax
  \@arrayskip\@preamble}
\makeatother

\begin{document}

\begin{titlepage}

\baselineskip=12pt

\rightline{OUTP-96-42P}
\rightline{TPI-MINN-96/10-T}
\rightline{hep-th/9607037}
\rightline{   }
\rightline{\today}

\vskip 0.6truein
\begin{center}

\baselineskip=24pt
{\Large\bf Conformal Dimensions from Topologically Massive Quantum Field
Theory}\\
\baselineskip=12pt
\vskip 0.8truein
{\bf G. Amelino-Camelia}$^a$, {\bf I. I. Kogan}$^{a,b}$ and {\bf R. J.
Szabo}$^a$\\

\vskip 0.3truein

$^a$ {\it Department of Theoretical Physics, University of Oxford\\ 1 Keble
Road, Oxford OX1 3NP, U.K.}\\

\bigskip

$^b$ {\it Theoretical Physics Institute\\ Physics Department, University of
Minnesota\\ 116 Church St. S.E., Minneapolis, MN 55455, U.S.A.}\\

\vskip 1.3 truein

\end{center}

\begin{abstract}

\baselineskip=12pt

We discuss the evaluation of observables in two-dimensional conformal field
theory using the topological membrane description. We show that the spectrum of
anomalous dimensions can be obtained perturbatively from the topologically
massive quantum field theories by computing radiative corrections to
Aharonov-Bohm scattering amplitudes for dynamical charged matter fields. The
one-loop corrections in the case of topologically massive Yang-Mills theory are
shown to coincide with the scaling dimensions of the induced ordinary and
supersymmetric WZNW models. We examine the effects of the dressing of a
topologically massive gauge theory by topologically massive gravity and show
that the one-loop contributions to the Aharonov-Bohm amplitudes coincide with
the leading orders of the KPZ scaling relations for two-dimensional quantum
gravity. Some general features of the description of conformal field theories
via perturbative techniques in the three-dimensional approach are also
discussed.

\end{abstract}

\end{titlepage}

\clearpage\newpage

\baselineskip=18pt

\section{Introduction}

Conformal field theories in two dimensions \cite{CFT} have many important
applications in theoretical physics, most notably to string theory \cite{books}
and critical phenomena in planar systems \cite{phase}. In this Paper, the
three-dimensional (``membrane'') description of conformal field theories will
be considered. It is known that two-dimensional conformal field theories can be
described in three-dimensional terms by using an amusing connection between
them and (2 + 1)-dimensional topological Chern-Simons gauge theories. This
connection was first discovered by Witten \cite{Wita}, who found an isomorphism
between the finite dimensional space of conformal blocks of the
Wess-Zumino-Novikov-Witten (WZNW) model with Kac-Moody symmetry $G_{\rm
L}\times G_{\rm R}$ defined on a Riemann surface $\Sigma$, and the finite
dimensional Hilbert space of Chern-Simons gauge theory with gauge group $G$
defined on a three-dimensional spacetime manifold ${\cal M} = \Sigma
\times\IR^{1}$. It was subsequently demonstrated in \cite{ms} using path
integral techniques that a Chern-Simons theory defined on a three-dimensional
manifold $\cal M$ with two-dimensional boundary $\partial{\cal M}$ induces the
chiral gauged WZNW model on the world-sheet $\partial{\cal M}$ (see \cite{ms'}
for further details). Different coset constructions can be obtained along the
same lines using several Chern-Simons theories and the three-dimensional
approach is a good starting point to tame the conformal zoo \cite{ms}.

This connection between apparently distinct quantum field theories provides
unexpected and intriguing relations between two- and three-dimensional physics.
The connection between a Chern-Simons gauge theory in three dimensions and a
conformal field theory in two dimensions  was  used in \cite{kogan1} to suggest
 the topological membrane approach to string theory. The modern formulation of
string theory is  the quantum theory of two-dimensional conformal fields on
random surfaces \cite{books}. The basic idea of the topological membrane
approach is to fill in the string world sheet and view it as the boundary of a
three-manifold. The emphasis of this approach is on the world sheet (rather
than target space) properties of the induced string theory.

The induced WZNW model will arise even from the topologically massive gauge
theory \cite{siegel}--\cite{djt} defined by the action
\beq
S_{TMGT} = -\frac{1}{2e^2} \int_{\cal
M}d^3x~\tr~\sqrt{g}~g^{\mu\lambda}g^{\nu\rho}F_{\mu\nu}F_{\lambda\rho}+ S_{CS}
\label{tmgt}\eeq
where the topological action
\beq S_{CS} = \int_{\cal M} {k\over4\pi}~\tr\left(A\wedge dA +
{2\over3}A\wedge A\wedge A\right)=\int_{\cal
M}d^3x~\frac{k}{4\pi}\epsilon^{\mu\nu\lambda}~\tr\left(A_\mu\partial_\nu
A_\lambda+\frac{2}{3}A_\mu A_\nu A_\lambda\right)
\label{csaction}\eeq
is the integral of the parity-violating Chern-Simons 3-form over the
three-dimensional spacetime manifold $\cal M$ with metric $g_{\mu\nu}$ of
Minkowski signature. Here $A=A_\mu(x)dx^\mu=A_\mu^a(x)T^adx^\mu$ is a gauge
connection of the trivial vector bundle over $\cal M$, where the anti-Hermitian
generators $T^a$ of the compact gauge group $G$ are normalized as
\beq
\tr~T^aT^b=\frac{1}{2}\delta^{ab}~~~~~,
\label{tanorm}\eeq
and $F=dA+[A,A]/2$ is the curvature of $A$. Unlike the pure Chern-Simons theory
(\ref{csaction}), the action (\ref{tmgt}) does not define a topological field
theory because the Yang-Mills kinetic term $F^{2}$ for the gauge field depends
explicitly on the $3$-dimensional metric $g_{\mu\nu}$ and there are propagating
degrees of freedom (massive vector bosons) with topological mass
\beq
M=ke^2/4\pi
\eeq
which means that the Hilbert space of this quantum field theory is
infinite-dimensional. The presence of induced degrees of freedom on the
boundary $\partial{\cal M}$ depends solely on the Chern-Simons term in the
action and is due to the fact that this term is not invariant under local gauge
transformations $A \rightarrow A^g = g^{-1}Ag + g^{-1} dg$ with nontrivial
gauge functions on the boundary, i.e. $g|_{\partial{\cal M}} \neq 0$. When
$\pi_3(G)\neq0$, it is necessary to quantize the Chern-Simons coefficient $k$
so that $k\in\IZ$ \cite{djt} to ensure that the quantum theory is invariant
under large gauge transformations which are not connected to the identity. In
the induced two-dimensional WZNW model, the integer $k$ then represents the
level of the current algebra generated by the gauge group $G$.

The appearance of a metric tensor in (\ref{tmgt}) is very important because it
is the only way to connect gauge degrees of freedom and gravity. Recall that to
 construct a string theory there must be two ingredients. The first one is a
conformal field theory on the string world-sheet and the second is a
fluctuating (random) geometry. The latter ingredient can be described by the
three-dimensional Einstein gravity action
\beq
S_{E} =\kappa\int_{\cal M}d^3x~ \sqrt{g}~R
\label{e}\eeq
where $R$ is the curvature of the metric $g_{\mu\nu}$ and $\kappa=1/2\pi G$ is
the Planck mass ($G$ is the gravitational constant). The model (\ref{e}) can be
considered as a topological Chern-Simons theory with gauge group $ISO(2,1)$
\cite{Town}. It was shown in \cite{carkog1} that in this case an integration
over the three-dimensional metric in the bulk interior of $\cal M$ leads to an
integration over the moduli space of two-dimensional complex structures on the
boundary $\partial{\cal M}$, with the same integration measure that naturally
occurs in string theory (the Weil-Petersson measure). The
reparametrization-ghost contributions were obtained in \cite{carkog2} using the
explicit wavefunctions on moduli space. These wavefunctions were found in
\cite{carlip2} in the case of the torus $\Sigma = T^{2}$.

Besides the Einstein term it is also natural to incorporate the gravitational
Chern-Simons term into the gravity action \cite{carkog1,wittencentral,kogan2}.
Indeed, in both the pure gauge and gravitational cases one-loop radiative
corrections to the Yang-Mills and Einstein actions induce Chern-Simons terms in
three-dimensions. The quantum geometrodynamics of the topological membrane is
described by topologically massive gravity \cite{djt} which is the sum of the
Einstein and gravitational Chern-Simons terms
\beq
S_{TMG}  = {k'\over8\pi}\int_{\cal M}d^3x~ \epsilon^{\mu\nu\lambda}
 \left(R_{\mu\nu\alpha\beta}\omega_{\lambda}^{\;\alpha\beta} +
  {2\over3}\omega_{\mu\alpha}^{\;\;\beta}\omega_{\nu\beta}^{\;\;\gamma}
  \omega_{\lambda\gamma}^{\;\;\alpha}\right) + S_{E}
  \label{tmg}  \eeq
where $\omega_{\mu\alpha}^{\;\;\beta}$ is the Levi-Civita spin connection for
$g_{\mu\nu}$. The first term in (\ref{tmg}) can be regarded as a Chern-Simons
action for an $SO(2,1)$ gauge theory with connection $\omega$. The gravity
theory (\ref{tmg}) is no longer topological and there are now propagating
graviton degrees of freedom with topological mass
\beq
\mu=8\pi\kappa/k'
\label{gravmass}\eeq
In the connection with conformal field theory, the gravitational Chern-Simons
coefficient $k'$, which need not be quantized, is proportional to the
corresponding two-dimensional central charge $c$. It was shown in \cite{kogan2}
that the resulting  quantum mechanics on moduli space leads to a holomorphic
dependence of the wavefunctions on moduli and it was conjectured, based on the
similarity between the $1/k'$ expansion in topologically massive gravity and
the $1/c$ expansion in two-dimensional quantum gravity, that the model
(\ref{tmg}) should be related to the gravity sector in string theory, i.e. to
Liouville
theory \cite{books,pol2}. This conjecture was (formally) proved in
\cite{carlip3} at the path integral level where it was demonstrated that
topologically massive gravity defined on a (2 + 1)-dimensional manifold $\cal
M$ induces two-dimensional quantum gravity, i.e.
Liouville theory, on the boundary $\partial{\cal M}$. The two-dimensional
cosmological constant $\Lambda$, which determines the scale
in Liouville theory defined by the action
\beq
S_L = \int_{\partial{\cal M}}d^2z~ \left(
\partial_{\bar{z}}\phi\partial_{z}\phi
+ QR^{(2)}\phi + \Lambda \e^{\alpha\phi} \right)~~~~~,
\label{liouville}\eeq
is equal to the square of the topological graviton mass
\cite{kogan3} (see also \cite{ashworth})
\beq
\Lambda=\mu^2
\eeq

An important issue in the above formal correspondences is to what extent the
quantum field theory in the bulk $\cal M$ describes the quantum characteristics
of the induced conformal field theory on $\partial{\cal M}$. In this Paper we
will discuss how to connect the basic observables of two-dimensional conformal
field theory with those of the topologically massive theories using
perturbative techniques. More precisely, we shall show that the conformal spin
eigenvalues of primary operators in the WZNW and Liouville models arise from a
perturbative expansion of scattering amplitudes up to one-loop order in the
topologically massive field theories above coupled to charged matter fields.
The idea behind this relation is as follows. The observables of two-dimensional
conformal field
theory can be expressed in terms of the vertex operators
\beq
V(z, \bar{z})= V_{\rm L}(z)V_{\rm R}(\bar{z})
\label{vertexop}\eeq
where the subscripts L and R represent the holomorphic and anti-holomorphic
sectors of the world-sheet theory. The gauge- and topologically-invariant
operators of the topologically massive gauge theory (\ref{tmgt}) are determined
by the path-ordered Wilson line operators
\beq
W_{R}[C]=~\tr_{R}~P\exp\left(i\int_CA\right)
\label{wilsonline}\eeq
where $C$ is an oriented contour in the 3-manifold $\cal M$. The Wilson line
(\ref{wilsonline}) describes the holonomy that arises from adiabatic transport
of a charged particle, in a representation $R$ of the gauge group $G$ and with
world-line $C$, in the presence of the gauge field $A$. When $\cal M$ is the
filled cylinder depicted in Fig. 1.1 (a), its two boundaries describe the left-
and right-moving sectors of the string worldsheet \cite{leith}. The Wilson line
(\ref{wilsonline}) coincides on the left boundary with the holomorphic part
$V_{\rm L}(z)$ and on the right boundary with the anti-holomorphic part $V_{\rm
R}(\bar{z})$ of the vertex operator (\ref{vertexop}). Thus the correlation
functions in the two-dimensional conformal field theory are related to
correlation functions of the topologically massive gauge theory (\ref{tmgt}).
The interchange of two points $z_1$ and $z_2$ on the worldsheet is equivalent
to the linking of the corresponding Wilson lines (see Fig. 1.1 (b)).

\begin{figure}
\unitlength=0.90mm
\linethickness{0.4pt}
\begin{picture}(153.00,49.00)(0,10)
\small
\bezier{124}(20.00,25.00)(6.00,34.00)(15.00,45.00)
\bezier{48}(15.00,45.00)(20.00,48.00)(25.00,45.00)
\bezier{128}(25.00,45.00)(35.00,35.00)(20.00,25.00)
\put(20.00,25.00){\line(1,0){50}}
\put(20.00,46.50){\line(1,0){50}}
\bezier{124}(70.00,25.00)(56.00,34.00)(65.00,45.00)
\bezier{48}(65.00,45.00)(70.00,48.00)(75.00,45.00)
\bezier{128}(75.00,45.00)(85.00,35.00)(70.00,25.00)
\bezier{124}(110.00,25.00)(96.00,34.00)(105.00,45.00)
\bezier{48}(105.00,45.00)(110.00,48.00)(115.00,45.00)
\bezier{128}(115.00,45.00)(125.00,35.00)(110.00,25.00)
\put(110.00,25.00){\line(1,0){50}}
\put(110.00,46.50){\line(1,0){50}}
\bezier{124}(160.00,25.00)(146.00,34.00)(155.00,45.00)
\bezier{48}(155.00,45.00)(160.00,48.00)(165.00,45.00)
\bezier{128}(165.00,45.00)(175.00,35.00)(160.00,25.00)
\put(45.00,15.00){\makebox(0,0){(a)}}
\put(135.00,15.00){\makebox(0,0){(b)}}
\put(20.00,30.00){\circle*{1.50}}
\put(20.00,40.00){\circle*{1.50}}
\put(70.00,30.00){\circle*{1.50}}
\put(70.00,40.00){\circle*{1.50}}
\put(110.00,30.00){\circle*{1.50}}
\put(110.00,40.00){\circle*{1.50}}
\put(160.00,30.00){\circle*{1.50}}
\put(160.00,40.00){\circle*{1.50}}
\thicklines
\put(20.00,30.00){\line(1,0){50}}
\put(20.00,40.00){\line(1,0){50}}
\thinlines
\put(15.00,24.00){\makebox(0,0){$(\partial{\cal M})_{\rm L}$}}
\put(77.00,24.00){\makebox(0,0){$(\partial{\cal M})_{\rm R}$}}
\put(105.00,24.00){\makebox(0,0){$(\partial{\cal M})_{\rm L}$}}
\put(167.00,24.00){\makebox(0,0){$(\partial{\cal M})_{\rm R}$}}
\put(20.00,32.50){\makebox(0,0){$z_1$}}
\put(20.00,42.50){\makebox(0,0){$z_2$}}
\put(70.00,32.50){\makebox(0,0){$z_1$}}
\put(70.00,42.50){\makebox(0,0){$z_2$}}
\put(110.00,32.50){\makebox(0,0){$z_1$}}
\put(110.00,42.50){\makebox(0,0){$z_2$}}
\put(160.00,32.50){\makebox(0,0){$z_2$}}
\put(160.00,42.50){\makebox(0,0){$z_1$}}
\put(45.00,32.50){\makebox(0,0){$C_1$}}
\put(45.00,42.50){\makebox(0,0){$C_2$}}
\put(45.00,50.00){\makebox(0,0){$\cal M$}}
\put(135.00,50.00){\makebox(0,0){$\cal M$}}
\thicklines
\put(110.00,30.00){\line(1,0){15}}
\put(110.00,40.00){\line(1,0){15}}
\put(125.00,30.00){\line(2,1){20}}
\put(125.00,40.00){\line(2,-1){8}}
\put(145.00,30.00){\line(-2,1){8}}
\put(145.00,30.00){\line(1,0){15}}
\put(145.00,40.00){\line(1,0){15}}
\thinlines
\put(128.00,28.00){\makebox(0,0){$C_1$}}
\put(128.00,42.00){\makebox(0,0){$C_2$}}
\end{picture}
\begin{description}
\small
\baselineskip=12pt
\item[Figure 1.1:] (a) The filled cylinder $\cal M$ whose boundaries
$(\partial{\cal M})_{\rm L,R}$ have punctures $z_i$ corresponding to insertions
of the vertex operators $V(z_i,\bar z_i)$. The punctures on the 2 boundaries
are joined together by curves $C_i\subset{\cal M}$ which represent the
propagation of charged particles in the bulk $\cal M$ whose holonomies are
described by the Wilson line operators $W_R[C_i]$. (b) An interchange of
punctures on the right-moving world-sheet $(\partial{\cal M})_{\rm R}$ is
equivalent to the linking of the corresponding Wilson lines in the bulk.
\end{description}
\end{figure}

In the simplest case of two-dimensional free bosons, i.e. a $c=1$ conformal
field theory, the chiral vertex operator is
\beq
V_{q}(z)=\e^{iq\phi(z)}
\label{vertexop1}\eeq
which can be obtained from the Wilson line in the corresponding
three-dimensional $U(1)$ theory (the representation $R$ in this case is
parametrized by an abelian charge $q$)
\beq
W_{q}[C]=\exp\left(iq\int_CA\right)
\label{wilsonline1}\eeq
with the conformal field $\phi$ as usual identified with the surviving
degree of freedom of the gauge field $A$ on $\partial{\cal M}$. It is known
that in this simple abelian case the short-distance operator product expansion
gives
\beq
\langle V_q(z_1)V_q(z_2)\rangle\sim(z_1-z_2)^{-q^2\Delta}
\label{OPE}\eeq
for $z_1\sim z_2$, where $\Delta$ is the scaling dimension of the corresponding
chiral vertex operators. Thus the interchange operation depicted in Fig. 1.1
(b) induces a phase factor $\e^{2\pi iq^2\Delta}$ in the vacuum expectation
value $\langle V_q(z_1)V_q(z_2)\rangle$. A similar phase factor appears in the
topologically massive gauge theory correlator $\langle W_q[C_1]W_q[C_2]\rangle$
in the bulk $\cal M$ due to the linking of the two corresponding Wilson lines
(when the minimal distance between the 2 charged particles is much larger than
the Compton length of the topologically massive vector boson), where $\Delta$
is the transmuted spin of the charged particles due to their interaction with
the Chern-Simons gauge field. This latter phase factor is manifested in the
Aharonov-Bohm part of the charged particle-particle scattering amplitudes in
the gauge theory \cite{kogmor}. Based on the above correspondences, we expect
the phase $\Delta$ to be the same in the two models. In the non-abelian case
these heuristic arguments are not precise, but we still expect the same sort of
relations between observables of the two- and three-dimensional quantum field
theories. A gravitational dressing of the above theories by the topologically
massive or Liouville gravity models would correspond to graviton exchanges
between the Wilson lines depicted in Fig. 1.1. In this Paper we shall
demonstrate the equivalences between the anomalous spins in the two- and
three-dimensional cases.

In the following, we shall compute one-loop radiative corrections to the
Aharonov-Bohm amplitudes for charged particle scattering in the topologically
massive gauge and gravity theories. We shall show that the anomalous dimensions
for the WZNW model \cite{kz} appear in this way from the non-abelian
topologically massive gauge theory (\ref{tmgt}) coupled to dynamical charged
scalar fields. For supersymmetric topologically massive Yang-Mills theory, we
get the anomalous dimensions for the $N=1$ supersymmetric WZNW model
\cite{divec}. Then, we shall study the effects of gravitational dressing on the
Aharonov-Bohm amplitudes and demonstrate that the one-graviton exchange
corrections to Aharonov-Bohm scattering coincide with the leading correction to
the bare conformal dimension predicted by the celebrated
Knizhnik-Polyakov-Zamolodchikov (KPZ) formula \cite{pol2}. In this way, the
topological membrane approach suggests a geometric origin for the KPZ scaling
relations.

The organization of this Paper is as follows. In Section 2 we briefly discuss
the structure of the Aharonov-Bohm amplitudes in Chern-Simons theory and how
they can be used to perturbatively determine anomalous dimensions of the
associated conformal field theories. In Section 3 we go through the standard
calculation of the one-loop renormalization of the Chern-Simons coefficient $k$
in the topologically massive gauge theory (\ref{tmgt}) coupled to charged
matter fields and show that the corresponding Aharonov-Bohm amplitude leads to
the anticipated spin. There we find that the correct dimension is largely
determined by using an appropriate Slavnov-Taylor identity for the gauge field
theory. In Section 4 we then include the coupling of the topologically massive
gauge theory to topologically massive gravity. There we demonstrate that the
standard set of Ward-Takahashi identities in the presence of the gravitational
dressing still apply and we use them to evaluate a large set of Feynman
diagrams contributing to the full scattering amplitude. The remaining graviton
exchange diagrams are then evaluated explicitly and we discuss the structure of
them and how they reproduce anticipated features of the conformal field theory.
Section 5 contains some concluding remarks and two Appendices at the end of the
Paper are devoted to technical details of the evaluation of some of the more
complicated Feynman graphs.

\section{Aharonov-Bohm Amplitudes and Anomalous Spin in Conformal Field Theory}

In this Section we shall describe the general form of the Aharonov-Bohm
amplitude in Chern-Simons theory and how it can be used for a perturbative
evaluation of conformal dimensions in the topologically massive field theories.
We shall assume here and in the following that the Chern-Simons coefficients
$k$ and $k'$ are large enough so that the perturbative expansions of the
respective quantum field theories make sense. We consider the minimal coupling
of the Chern-Simons gauge field $A$ to a conserved current $J^\mu=J_a^\mu R^a$,
with $R(G)$ some irreducible unitary representation of $G$ whose generators are
normalized as
\beq
{\rm tr}~R^aR^b=T_R(G)\delta^{ab}~~~~~,
\label{trdef}\eeq
by adding the term
\beq
S_J=\int d^3x~2~{\rm tr}~J^\mu A_\mu^aR^a
\label{mattercoupling}\eeq
to the Chern-Simons action (\ref{csaction}). From (\ref{csaction}) it follows
that the momentum space bare gluon propagator in the transverse Landau gauge
(i.e. see the next Section) is
\beq
{\cal G}^{ab}_{\mu\nu}(p)=\left\langle
A_\mu^a(p)A_\nu^b(-p)\right\rangle_A=-\frac{4\pi}{k}
\delta^{ab}\frac{\epsilon_{\mu\nu\lambda}p^\lambda}{p^2}
\label{puregaugeprop}\eeq
where the expectation value is taken in the non-interacting part of the gauge
theory (\ref{csaction}). The invariant amplitude for the scattering of two
charged particles of initial momenta $p_1$ and $p_2$ represented by the current
$J$ is then of the form
\beq
{\cal A}(p_1,p_2;q)\equiv i~\tr~J^\mu(2p_1-q){\cal
G}_{\mu\nu}(q)J^\nu(2p_2+q)=-\frac{16\pi
i}{k}\dim(G)T_R(G)f_G(k)\frac{\epsilon_{\mu\nu\lambda}p_1^\mu p_2^\nu
q^\lambda}{q^2}
\label{abrelgen}\eeq
where $q$ is the momentum transfer and we have used the transversality of the
gauge propagator $p^\mu{\cal G}_{\mu\nu}(p)=0$ and current conservation $p_\mu
J^\mu(p)=0$. Here $f_G(k)=\sum_{n\geq0}f_n/k^n$ is some function whose
coefficients $f_n$ depend only on invariants of the gauge group $G$ and which
can be computed perturbatively order by order in the Chern-Simons coupling
constant $1/k$.

In the center of momentum frame the amplitude (\ref{abrelgen}) becomes
\beq
{\cal A}(p_1,p_2;q)=\frac{16\pi i}{k}\dim(G)T_R(G)f_G(k)\frac{\vec p\times\vec
q}{\vec q^2}
\label{abnonrelgen}\eeq
This amplitude is none other than the Aharonov-Bohm amplitude for the
scattering of a charge of strength $\sqrt{T_R(G)f_G(k)}$ off of a flux of
strength $(4\pi/k)\dim(G)\sqrt{T_R(G)f_G(k)}$. This is the standard argument
for the appearence of induced fractional spin and statistics perturbatively in
a Chern-Simons gauge theory and it leads to the spin factor
\beq
\Delta_G(k)=\frac{T_R(G)}{k}f_G(k)
\label{spin}\eeq
which measures the anomalous change of phase in the Aharonov-Bohm wavefunction
under adiabatical rotation of one charged particle about another in the gauge
theory (\ref{csaction}) \cite{kogmor}. Thus computing the function $f_G(k)$
order by order in perturbation theory allows us to check the standard formulas
for anomalous spin in the Chern-Simons field theories.

There are two well-known conformal dimension formulas that can be explored
perturbatively in this way. It was shown by Knizhnik and Zamolodchikov
\cite{kz} that the conformal weights of the primary operators in a current
algebra based on a compact semi-simple Lie group $G$ are given by
\beq
\Delta_{KZ}=\frac{T_R(G)}{k+C_2(G)}
\label{kzweights}\eeq
where $k\in\IZ$ is the level of the WZNW model, $R(G)$ is an irreducible
unitary representation of $G$ carried by the primary conformal fields, and
$C_2(G)$ is the quadratic Casimir operator in the adjoint representation of
$G$, i.e.
\beq
C_2(G)\delta^{ab}=f^{acd}f^{bcd}
\label{c2def}\eeq
with $f^{abc}$ the antisymmetric structure constants of the Lie group $G$,
\beq
[T^a,T^b]=f^{abc}T^c
\label{strucdef}\eeq
The perturbative expansion (\ref{spin}) can be therefore compared with the
large-$k$ expansion of the weights (\ref{kzweights}) for the WZNW model which
is
\beq
\Delta_{KZ}=\frac{T_R(G)}{k}\left[1-\frac{C_2(G)}{k}+\left(\frac{C_2(G)}{k}
\right)^2-\dots\right]
\label{KZexp}\eeq

In the case of the 2-dimensional Liouville theory, it is known from the KPZ
random geometry approach \cite{pol2} that the bare conformal dimensions
$\Delta_0$ of primary fields undergo a simple quadratic transformation
\beq
\Delta-\Delta_0=\frac{\Delta(1-\Delta)}{\tilde k+2}
\label{KPZtransf}\eeq
as a result of the gravitational dressing. Here $\tilde k$ is the central
charge of the $SL(2,\IR)$ current algebra. It is related to the gravitational
Chern-Simons coefficient in (\ref{tmg}) by $\tilde k=-k'-4$ (the coefficient of
the Liouville kinetic term in (\ref{liouville}) is $k'$ after an appropriate
rescaling of the conformal field $\phi$). The solution of (\ref{KPZtransf}) for
the gravitationally-dressed weight $\Delta$ with the branch satisfying the
``natural" boundary condition $\Delta(\Delta_0=0)=0$ is
\beq
\Delta=-\frac{(k'+1)}{2}\left(1-\sqrt{1+\frac{4\Delta_0(k'+2)}
{(k'+1)^2}}\right)
\label{KPZwt}\eeq
whose expansion for large $k'$ is
\beq
\Delta=\Delta_0+\frac{\Delta_0-\Delta_0^2}{k'}+\frac{2\Delta_0^3-\Delta_0-
\Delta_0^2}{k'^2}+\dots
\label{KPZexp}\eeq
Thus, coupling the gravitational theory to charged matter of conformal weight
$\Delta_0$, one can also compute the gravitational corrections to the
Aharonov-Bohm amplitudes above perturbatively in $1/k'$ and check it with
(\ref{KPZexp}).

\section{Conformal Weights in Topologically Massive Gauge Theory}

In this Section we shall demonstrate, as a preliminary discussion, the
appearence of the Knizhnik-Zamolodchikov formula perturbatively up to one-loop
order in Chern-Simons gauge theory. This calculation is essentially the
well-known computation of the one-loop corrections to the Chern-Simons
coefficient $k$ and we shall therefore only highlight the details of the
calculation for the sake of comparison with the corresponding calculations that
we will present later on for the gravitational contributions. We assume here
that $\cal M$ is flat Minkowski spacetime with metric $g=~{\rm diag}(1,-1,-1)$.
In the next Section we will remove this constraint and consider the effects of
gravitational dressing. In the previous Section we saw that scattering
amplitudes in the Chern-Simons gauge theory correspond to the familiar ones
that appear from the Aharonov-Bohm effect. However, the Chern-Simons gauge
theory (\ref{csaction}) is only strictly-renormalizable and the quantum field
theory should be properly defined with an ultraviolet cutoff. We therefore add
a non-topological parity-symmetric Yang-Mills kinetic term to the action and
consider the topologically massive gauge theory (\ref{tmgt}) instead. The
dimensionful parameter $e^2$ acts as an ultraviolet cutoff, and the effective
coupling constant of the perturbation expansion is $e^2/M\sim1/k$. The
appearence of a massive, propagating gluon in the model removes the infrared
divergences and makes the quantum field theory with action (\ref{tmgt}) a
power-counting super-renormalizable field theory with finite, computable
physical parameters \cite{prao}--\cite{cswu}. The Aharonov-Bohm amplitudes
(\ref{abrelgen}) can then be obtained from the imaginary, parity-odd parts of
the amplitudes in the topologically massive gauge theory in the infrared limit
$e^2\to\infty$ when the Yang-Mills action in (\ref{tmgt}) becomes irrelevant
and the resulting topological field theory induces the WZNW model on
$\partial{\cal M}$. This limit, wherein the only effect of the gluon field is
to transmute the spin and statistics of external charged particles, is often
refered to as the `anyon limit'\footnote{\baselineskip=12pt Note that in the
anyon limit the system is essentially non-relativistic. In fact, the
Aharonov-Bohm scattering can be investigated using the formalism of
non-relativistic quantum field theory. There, however, the role of the
ultraviolet regulator (played in the present relativistic formalism by the
Yang-Mills kinetic term) can be played by a quartic contact interaction among
the matter fields \cite{abdelta}.}.

We use the standard Faddeev-Popov gauge-fixing procedure to fix a covariant
gauge by introducing anticommuting ghost fields $\eta^a,\bar\eta^a$ in the
adjoint representation of $G$ and adding the action
\beq
S_g=\int d^3x~\left(\frac{1}{\xi}~\tr(\partial_\mu
A^\mu)^2+(\partial^\mu\bar\eta^a)(\partial_\mu\eta^a+if^{abc}A_\mu^b\eta^c)
\right)
\label{gaugeaction}\eeq
to (\ref{tmgt}), where $\xi$ is the covariant gauge fixing parameter and the
ghost field action in (\ref{gaugeaction}) is minimally coupled to the gauge
field $A$. It is then straightforward to write down the Feynman rules for this
gauge theory. The bare gluon propagator (\ref{puregaugeprop}) is replaced by
\beq
G_{\mu\nu}^{ab}(p)=\delta^{ab}\left\{-ie^2\left(\frac{p^2g_{\mu\nu}^
\perp(p)+iM
\epsilon_{\mu\nu\lambda}p^\lambda}{p^2(p^2-M^2)}\right)+\xi\frac{p_\mu
p_\nu}{(p^2)^2}\right\}
\label{tmgaugeprop}\eeq
where
\beq
g_{\mu\nu}^\perp(p)=g_{\mu\nu}-p_\mu p_\nu/p^2
\eeq
is the symmetric transverse projection operator in momentum space with $p^\mu
g_{\mu\nu}^\perp(p)=0$. From (\ref{gaugeaction}) it follows that the bare ghost
propagator in momentum space is
\beq
\tilde G^{ab}(p)=\langle\bar\eta^a(p)\eta^b(-p)\rangle_{\eta}=i\delta^{ab}/p^2
\label{ghostprop}\eeq
The bare ghost-ghost-gluon vertex function is
\beq
\tilde\Gamma_\mu^{abc}(p,q;r)=-if^{abc}p_\mu
\label{ghghgl}\eeq
where $r=p+q$, the bare 3-point gluon vertex function is
\beq
\Gamma_{\mu\nu\lambda}^{abc}(p,q,r)=\frac{i}{e^2}f^{abc}\left(iM
\epsilon_{\mu\nu
\lambda}+(p-q)_\lambda g_{\mu\nu}+(q-r)_\mu g_{\nu\lambda}+(r-p)_\nu
g_{\lambda\mu}\right)
\label{glglgl}\eeq
where $p+q+r=0$, and the bare 4-point gluon vertex is given by
\beq\new{\begin{array}{c}
\Gamma_{\mu\nu\sigma\lambda}^{abcd}(p,q,r,s)=\frac{i}{e^2}\left[f^{eab}f^{ecd}
\left(g_{\mu\lambda}g_{\nu\sigma}-g_{\mu\sigma}g_{\nu\lambda}\right)+f^{eac}
f^{edb}\left(g_{\mu\sigma}g_{\lambda\nu}-g_{\mu\nu}g_{\lambda\sigma}\right)
\right.\\\left.+f^{ead}f^{ebc}\left(g_{\mu\nu}g_{\sigma\lambda}-g_{\mu
\lambda}g_{\sigma\nu}\right)\right]\end{array}}
\label{4gl}\eeq
with $p+q+r+s=0$.

\subsection{Coupling to Scalar Fields}

We first consider the minimal coupling of the gauge theory defined above to
dynamical charged massive scalar fields with action
\beq
S_m=\int d^3x~\left(\left|\partial_\mu\phi_A-iR_{AB}^aA_\mu^a\phi_B\right|^2
-m^2|\phi_A|^2\right)
\label{scalaraction}\eeq
where $R(G)$ is the representation of $G$ carried by the charged mesons and $m$
is their mass. Then the bare scalar propagator is
\beq
S_{BA}(p)=\left\langle\phi^*_A(p)\phi_B(-p)\right\rangle_\phi
=\frac{i}{p^2-m^2}\delta_{AB}
\label{scalarprop}\eeq
and the bare meson-meson-gluon vertex function is $-iR_{AB}^a(p+p')_\mu$ (with
$q=p-p'$ the gluon momentum) while the meson-meson-gluon-gluon vertex is
$ig_{\mu\nu}\{R^a,R^b\}_{AB}$. To evaluate the one-loop order conformal weight
$\Delta^{(1)}$ in this topologically massive gauge theory, we consider the
Feynman graphs corresponding to the charged meson-meson scattering amplitude up
to one-loop order. We calculate all amplitudes in the following in the
transverse Landau gauge ($\xi=0$ in (\ref{tmgaugeprop})) and we shall always
assume that the external scalar particles are on-shell. The tree-level
amplitude is easily calculated to be (Fig. 3.1)
\beq\new{\begin{array}{lll}
{\cal T}(p_1,p_2;q)&=&-i~{\rm tr}~R^a(2p_1-q)^\mu G_{\mu\nu}^{ab}(q)
R^b(2p_2+q)^\nu\\&=&-\frac{4e^2\dim(G)T_R(G)}{q^2(q^2-M^2)}\left\{\left[q^2
(p_1\cdot p_2)-(p_1\cdot q)(p_2\cdot
q)\right]+iM\epsilon_{\mu\nu\lambda}p_1^\mu p_2^\nu
q^\lambda\right\}\end{array}}
\label{treerel}\eeq
which in the anyon limit is
\beq
\lim_{M\to\infty}{\cal T}(p_1,p_2;q)=\frac{16\pi
i\dim(G)T_R(G)}{k}\frac{\epsilon_{
\mu\nu\lambda}p_1^\mu p_2^\nu q^\lambda}{q^2}
\label{treeab}\eeq
This identifies the tree-level conformal weight
\beq
\Delta^{(0)}=T_R(G)/k
\label{wttree}\eeq
which agrees with the leading order term in the expansion (\ref{KZexp}) in
$1/k$.

\begin{figure}
\begin{picture}(30000,10000)
\small
\put(21000,3500){\makebox(0,0){$q$}}
\put(14000,1000){\makebox(0,0){$p_2$}}
\put(27000,1000){\makebox(0,0){$p_2+q$}}
\put(14000,6000){\makebox(0,0){$p_1$}}
\put(27000,6000){\makebox(0,0){$p_1-q$}}
\drawline\fermion[\E\REG](20000,1000)[5000]
\drawline\fermion[\W\REG](20000,1000)[5000]
\drawline\photon[\N\REG](20000,1000)[5]
\drawline\fermion[\E\REG](\photonbackx,\photonbacky)[5000]
\drawline\fermion[\W\REG](\photonbackx,\photonbacky)[5000]
\end{picture}
\begin{description}
\small
\baselineskip=12pt
\item[Figure 3.1:] Tree-level amplitude in topologically massive Yang-Mills
theory. The conventions to be used for all Feynman diagrams in the following
are that $p_1$, $p_2$ denote the incoming matter momenta, and $q$ is the
momentum transfer. Straight lines denote matter fields while wavy lines depict
gluons.
\end{description}
\end{figure}

Next we evaluate the corrections to the internal gluon line in Fig. 3.1 from
the gluon, ghost and scalar loops, and the gluon and scalar tadpoles (see Fig.
3.2). These yield the amplitude
\beq
{\cal G}(p_1,p_2;q)=-i~{\rm tr}~R^a(2p_1-q)^\mu R^b(2p_2+q)^\nu G_{\mu\lambda}
^{ac}(q)\Pi^{\lambda\rho}_{cd}(q)G^{db}_{\rho\nu}(q)
\label{vacpoltot}\eeq
where $\Pi_{\mu\nu}^{ab}(q)$ is the total 1-loop vacuum polarization tensor
which renormalizes the bare gluon propagator as $G_{\mu\nu}^{ab}(q)^{-1}\to
G_{\mu\nu}^{ab}(q)^{-1}+\Pi_{\mu\nu}^{ab}(q)$. Since $p^\mu p^\nu
G_{\mu\nu}^{ab}(q)=\xi\delta^{ab}$, by gauge invariance it follows that the
longitudinal part is not renormalized and the gluon self energy is transverse
in $q$,
\beq
\Pi_{\mu\nu}^{ab}(q)=\delta^{ab}\left\{\frac{q^2g_{\mu\nu}^\perp(q)}{M}
\Pi_e(q^2)-\frac{k}{4\pi}\Pi_o(q^2)\epsilon_{\mu\nu\lambda}q^\lambda\right\}
\label{gselftot}\eeq
This result follows from the Ward-Takahashi identities of the gauge theory.
As pointed out in \cite{kogsem}, radiative corrections from the scalar loop
induce a finite renormalization of the charge parameter $e^2$ which leads to a
renormalized charge $e_r^2(e^2)$ which is finite as $e^2\to\infty$. It is
therefore more natural to consider not the anyon limit of the topologically
massive gauge theory, but the equivalent limit $q\to0$ of small momentum
transfer. As has been discussed extensively in \cite{prao}--\cite{cswu}, the
topologically massive gauge theory is infrared finite in the Landau gauge and
so all renormalized quantities are non-singular at $q=0$. In fact, at least to
2-loop order, all renormalization constants are independent of momenta and
coincide exactly with their zero momentum limit. Substituting (\ref{gselftot})
into (\ref{vacpoltot}) and taking this limit, its imaginary part
becomes\footnote{\baselineskip=12pt Radiative corrections also introduce a
finite contribution in the anyon limit to the real part of the amplitude which
can be attributed to the finite renormalization of the charge parameter $e^2$
and a short-ranged Pauli interaction between the magnetic fluxes carried by the
anyons that are induced by the Chern-Simons gauge field
\cite{kogsem,szkogsem}.}
\beq
\lim_{q\to0}~{\rm Im}~{\cal
G}(p_1,p_2;q)=-\frac{16\pi\dim(G)T_R(G)}{k}~\Pi_o(0)~
\frac{\epsilon_{\mu\nu\lambda}p_1^\mu p_2^\nu q^\lambda}{q^2}
\label{ganyon}\eeq
which leads to the one-loop vacuum polarization contribution
\beq
\Delta_{\rm vac}^{(1)}=-\frac{T_R(G)}{k}~\Pi_o(0)
\label{wtvac}\eeq
to the induced spin of the scalar fields.

\begin{figure}
\begin{picture}(30000,10000)
\small
\drawline\fermion[\E\REG](20000,0)[5000]
\drawline\fermion[\W\REG](20000,0)[5000]
\drawline\photon[\N\REG](20000,0)[2]
\thicklines
\put(20000,3000){\circle{2000}}
\thinlines
\put(20000,3000){\makebox(0,0){$\Pi$}}
\drawline\photon[\N\REG](20000,4000)[2]
\drawline\fermion[\E\REG](\photonbackx,\photonbacky)[5000]
\drawline\fermion[\W\REG](\photonbackx,\photonbacky)[5000]
\end{picture}
\begin{center}
\begin{picture}(40000,10000)
\small
\put(1000,1000){\makebox(0,0){$\mu,a$}}
\put(0,0){\makebox(0,0){$q$}}
\put(1000,7000){\makebox(0,0){$\nu,b$}}
\put(0,8000){\makebox(0,0){$q$}}
\drawline\photon[\N\REG](0,1000)[2]
\thicklines
\put(0,4000){\circle{2000}}
\thinlines
\put(0,4000){\makebox(0,0){$\Pi$}}
\drawline\photon[\N\REG](0,5000)[2]
\put(3000,4000){\makebox(0,0){$=$}}
\drawline\photon[\N\REG](6000,1000)[2]
\put(6000,4000){\circle{2000}}
\drawline\photon[\N\REG](6000,5000)[2]
\put(9000,4000){\makebox(0,0){$+$}}
\drawline\photon[\N\REG](12000,1000)[6]
\drawline\photon[\NE\FLIPPED](12000,2200)[3]
\drawline\photon[\SE\REG](12000,6000)[3]
\put(17000,4000){\makebox(0,0){$+$}}
\drawline\photon[\N\REG](20000,1000)[2]
\multiput(20000,3000)(200,200){6}{\line(300,0){200}}
\multiput(20000,3000)(-200,200){6}{\line(300,0){200}}
\multiput(20000,5000)(200,-200){6}{\line(300,0){200}}
\multiput(20000,5000)(-200,-200){6}{\line(300,0){200}}
\drawline\photon[\N\REG](20000,5000)[2]
\put(23000,4000){\makebox(0,0){$+$}}
\drawline\photon[\N\REG](26000,1000)[6]
\put(27150,4000){\circle{2000}}
\put(31000,4000){\makebox(0,0){$+$}}
\drawline\photon[\N\REG](33000,1000)[6]
\drawline\photon[\NE\FLIPPED](33000,4000)[2]
\drawline\photon[\SE\REG](33000,4000)[2]
\drawline\photon[\SE\REG](34250,5250)[2]
\drawline\photon[\NE\FLIPPED](34250,2750)[2]
\put(38000,4000){\makebox(0,0){$+~~\dots$}}
\end{picture}
\end{center}
\begin{description}
\small
\baselineskip=12pt
\item[Figure 3.2:] Total one-loop vacuum polarization contribution to the
amplitude. The gluon self-energy is $\Pi_{\mu\nu}^{ab}(q)$, and the dashed
lines denote the Faddeev-Popov ghost fields.
\end{description}
\end{figure}

Consider first the contribution from the scalar loop and the scalar tadpole.
They are given, respectively, by
\beq
\Pi_{\mu\nu}^{(s)ab}(q)=-\int\frac{d^3k}{(2\pi)^3}~\left\{{\rm tr}~(2k-q)_\mu
R^a(2k-q)_\nu R^bS(k)S(k-q)+2ig_{\mu\nu}~\tr~R^aR^bS(k)\right\}
\label{pis}\eeq
which can be evaluated in the infrared limit to give \cite{sswu}
\beq
\lim_{q\to0}\Pi_{\mu\nu}^{(s)ab}(q)=-i\delta^{ab}\frac{T_R(G)}{24\pi
m}q^2g_{\mu\nu}^\perp(q)
\label{pis0}\eeq
The correction (\ref{pis0}) renormalizes the charge parameter to
$1/e_r^2=1/e^2+T_R(G)/24\pi m$ which has the finite value
$e_r^2(e^2=\infty)=24\pi m/T_R(G)$ in the anyon limit, as mentioned
above\footnote{\baselineskip=12pt Strictly speaking, we should take the limit
$m\to\infty$ where the mesons become fixed external sources. As we shall see,
the associated conformal weights are independent of any mass parameters of the
model and thus in this limit we would obtain the same results. The anyon limit
can then be reached in this regime even in the renormalized theory.}. The
polarization tensor (\ref{pis}) has only a parity-even part,
$\Pi_o^{(s)}(q^2)=0$, which owes to the well-known fact that the scalar loop
does not renormalize the Chern-Simons coefficient $k$ \cite{prao,sswu}. It
therefore does not contribute to the Aharonov-Bohm part of the amplitude
(\ref{vacpoltot}). Next, consider the contributions from the gluon and ghost
loops, and the gluon tadpole diagram, respectively,
\beq\new{\begin{array}{lll}
\Pi_{\mu\nu}^{(g)ab}(q)&=&\int\frac{d^3k}{(2\pi)^3}~\left\{\Gamma^{afe}
_{\mu\lambda
\alpha}(-q,k,q-k)\Gamma^{bcd}_{\nu\rho\beta}(q,-k,k-q)G^{\lambda\rho}_{fc}(k)
G_{de}^{\beta\alpha}(k-q)\right.\\&
&\left.+\tilde\Gamma^{efa}_\mu(k-q,k;q)\tilde\Gamma^{cdb}
_\nu(k,k-q;q)\tilde G^{fc}(k)\tilde
G^{de}(k-q)+\Gamma^{abcd}_{\mu\nu\lambda\rho}(q,q,k,k)G_{cd}^{\lambda\rho}(k)
\right\}\end{array}}
\label{iloopglvac}\eeq
The only parity-odd contribution comes from the parity-odd part of the first
term in (\ref{iloopglvac}) (from the gluon loop). To find it, we contract
(\ref{iloopglvac}) with
$\frac{k}{4\pi}\frac{\epsilon_{\mu\nu\lambda}q^\lambda}{2q^2}$, and after some
algebra we get
\beq
\Pi_o^{(g)}(q^2)=-\frac{4\pi}{k}\frac{C_2(G)}{q^2}M\int\frac{d^3k}{(2\pi)^3}~
\frac{[k^2q^2-(k\cdot q)^2][5k^2+5(k\cdot
q)+4q^2+2M^2]}{k^2(k^2-M^2)(k+q)^2[(k+q)^2-M^2]}
\label{piogl}\eeq
The integral in (\ref{piogl}) is convergent and can be evaluated explicitly
using standard methods \cite{prao,cswu,gmrr}. Doing so, and then taking the
low-energy limit we find
\beq
\Pi_o(0)=\frac{7}{3k}C_2(G)~{\rm sgn}(k)
\label{pioddlowenergy}\eeq
which gives
\beq
\Delta^{(1)}_{\rm vac}=-\frac{7}{3k^2}T_R(G)C_2(G)~{\rm sgn}(k)
\label{vactot}\eeq

Next, we consider the 1-loop corrections to the meson-meson-gluon vertex (Fig.
3.3), which lead to the amplitude
\beq{\cal V}(p_1,p_2;q)=~{\rm tr}~R^a(2p_1-q)^\mu R^b\Gamma^\nu(p_2,-q)
G_{\mu\nu}^{ab}(q)+~{\rm tr}~R^a\Gamma^\mu(p_1,q)R^b(2p_2+q)^\nu
G_{\mu\nu}^{ab}(q)
\label{vertex}\eeq
where $\Gamma_\mu(p,q)$ is the irreducible vertex function which by gauge
invariance is given in terms of renormalization constants as
\beq
\Gamma_\mu(p,q)=-i(Z_e-1)(2p-q)_\mu-\frac{2i}{m}(Z_o-1)\epsilon_{\mu\nu\lambda}
p^\nu q^\lambda
\label{vertexfn}\eeq
Here $Z_e$ is the meson charge renormalization constant. The only relevant
contributions in the amplitude (\ref{vertex}) will be from the longitudinal
part of (\ref{vertexfn}) contracted with the partity-odd part of the gauge
propagator, or else from a parity-odd part of (\ref{vertexfn}) which is
singular at $q=0$ contracted with the parity-even part of the gauge propagator.
By the infrared finiteness of the topologically massive gauge theory in the
Landau gauge, this latter possibility does not occur. This can also be checked
explicitly by examining the infrared structure of the Feynman integrals
involved. For instance, it was shown in \cite{kogsem} that the first, triangle
graph in Fig. 3.3 gives $Z_o^{(1)}=1/k$ at one-loop order which is just the
induced fractional spin of the anyons leading to an anomalous Pauli magnetic
moment interaction. This piece is therefore non-singular at $q=0$ and does not
contribute to the Aharonov-Bohm scattering. The former, longitudinal
contribution is easily found by exploiting the Slavnov-Taylor identities for
the non-abelian gauge theory which express the gauge invariance of the
renormalized quantum field theory through the universality of the gauge
coupling. The relation we shall use relates the meson-gauge 3-point coupling to
the ghost-gauge coupling.

\begin{figure}
\begin{center}
\begin{picture}(20000,8000)
\small
\drawline\fermion[\E\REG](1000,0)[5000]
\drawline\fermion[\W\REG](1000,0)[5000]
\drawline\photon[\N\REG](1000,0)[3]
\thicklines
\put(1000,4000){\circle{2000}}
\thinlines
\put(1000,4000){\makebox(0,0){$\Gamma$}}
\drawline\fermion[\E\REG](2000,4000)[4000]
\drawline\fermion[\W\REG](0,4000)[4000]
\put(10000,1500){\makebox{$+$}}
\drawline\fermion[\W\REG](18000,0)[4000]
\drawline\fermion[\E\REG](20000,0)[4000]
\thicklines
\put(19000,0){\circle{2000}}
\thinlines
\put(19000,0){\makebox(0,0){$\Gamma$}}
\drawline\photon[\N\REG](19000,1000)[3]
\drawline\fermion[\W\REG](\photonbackx,\photonbacky)[5000]
\drawline\fermion[\E\REG](\photonbackx,\photonbacky)[5000]
\end{picture}
\end{center}
\begin{center}
\begin{picture}(45000,8000)
\small
\drawline\fermion[\W\REG](2000,3000)[2000]
\drawline\fermion[\E\REG](4000,3000)[2000]
\drawline\photon[\N\REG](3000,0)[2]
\put(2000,0){\makebox(0,0){$q$}}
\put(4000,0){\makebox(0,0){$\mu$}}
\put(0,4000){\makebox(0,0){$p$}}
\put(6000,4000){\makebox(0,0){$p-q$}}
\thicklines
\put(3000,3000){\circle{2000}}
\thinlines
\put(3000,3000){\makebox(0,0){$\Gamma$}}
\put(9000,1500){\makebox(0,0){$=$}}
\drawline\photon[\N\REG](14000,0)[3]
\drawline\fermion[\E\REG](\photonbackx,\photonbacky)[3000]
\drawline\fermion[\W\REG](\photonbackx,\photonbacky)[3000]
\drawline\photon[\NE\REG](12200,3000)[3]
\drawline\photon[\NW\FLIPPED](16000,3000)[3]
\put(19000,1500){\makebox(0,0){$+$}}
\drawline\photon[\N\REG](25000,0)[3]
\drawline\fermion[\W\REG](\photonbackx,\photonbacky)[4000]
\drawline\fermion[\E\REG](\photonbackx,\photonbacky)[2000]
\drawline\photon[\NW\FLIPPED](25000,3000)[2]
\drawline\photon[\NE\REG](22500,3000)[2]
\put(29000,1500){\makebox(0,0){$+$}}
\drawline\photon[\N\REG](33000,0)[3]
\drawline\fermion[\W\REG](\photonbackx,\photonbacky)[2000]
\drawline\fermion[\E\REG](\photonbackx,\photonbacky)[4000]
\drawline\photon[\NW\FLIPPED](35500,3000)[2]
\drawline\photon[\NE\REG](33000,3000)[2]
\end{picture}
\end{center}
\begin{center}
\begin{picture}(20000,5000)
\small
\put(0,1500){\makebox(0,0){$+$}}
\drawline\photon[\N\REG](4000,0)[2]
\drawline\photon[\NE\FLIPPED](4000,2000)[2]
\drawline\fermion[\W\REG](\photonbackx,\photonbacky)[1000]
\drawline\fermion[\E\REG](\photonbackx,\photonbacky)[1000]
\drawline\photon[\NW\REG](4000,2000)[2]
\drawline\fermion[\W\REG](\photonbackx,\photonbacky)[1000]
\drawline\fermion[\E\REG](\photonbackx,\photonbacky)[1500]
\put(9000,1500){\makebox(0,0){$+$}}
\drawline\photon[\N\REG](14000,0)[2]
\drawline\photon[\NW\REG](14000,2000)[2]
\drawline\photon[\SW\FLIPPED](14000,4500)[2]
\drawline\fermion[\W\REG](14000,4500)[3000]
\drawline\fermion[\E\REG](14000,4500)[3000]
\drawline\photon[\NE\FLIPPED](14000,2000)[2]
\drawline\photon[\SE\REG](14000,4500)[2]
\put(21000,1500){\makebox(0,0){$+~~\dots$}}
\end{picture}
\end{center}
\begin{description}
\small
\baselineskip=12pt
\item[Figure 3.3:] Total one-loop vertex correction to the scattering
amplitude. The proper vertex function is $\Gamma_\mu(p,q)$.
\end{description}
\end{figure}

Consider the meson self energy operator $\Sigma(p)$ (Fig. 3.4) which
renormalizes the bare meson propagator as $S(p)^{-1}\to S(p)^{-1}+\Sigma(p)$.
The self-energy can be expressed in terms of the scalar wavefunction
renormalization constant $Z_\phi$ as
\beq
\Sigma(p)=-i(Z_\phi-1)(p^2-m^2)
\label{selfenergy}\eeq
Similarly, we introduce the ghost self-energy operator
$\tilde\Pi^{ab}(p)=\delta^{ab}\tilde\Pi(p)$ which renormalizes the bare ghost
field propagator as $\tilde G(p)^{-1}\to\tilde G(p)^{-1}+\tilde\Pi(p)$ and
which can be written in terms of the ghost wavefunction renormalization
constant $\tilde Z$ as
\beq
\tilde\Pi(p)=-ip^2(\tilde Z-1)
\label{ghostself}\eeq
Finally, we introduce the renormalized ghost-ghost-gluon vertex by
\beq
\tilde\Gamma_\mu^{(r)abc}(p)=-if^{abc}(\tilde Z_g-1)p_\mu
\label{gggvertexren}\eeq
where $\tilde Z_g$ is the ghost charge renormalization constant. The desired
Slavnov-Taylor identity is then
\beq
\frac{Z_\phi}{Z_e}=\frac{\tilde Z}{\tilde Z_g}
\label{slavtaylor}\eeq
and it expresses the fact that the mesons and ghosts couple in the same way to
the non-abelian gauge field $A$. As shown in \cite{prao,cswu}, $\tilde Z_g=1$
at least to two-loop order in perturbation theory. At one-loop order, we write
the renormalization constants as $Z=1+Z^{(1)}$, where $Z^{(n)}$ denotes the
order $1/k^n$ perturbative contribution. Then the Slavnov-Taylor identity
(\ref{slavtaylor}) at one-loop order is
\beq
Z_e^{(1)}=Z_\phi^{(1)}-\tilde Z^{(1)}
\label{stid1loop}\eeq
or, by the definitions of the renormalization constants, we have the identity
\beq
q^\mu\Gamma^{(1)}_\mu(p,q)+\Sigma^{(1)}(p-q)-\Sigma^{(1)}(p)=\tilde\Pi^{(1)}
(p)-\tilde\Pi^{(1)}(p-q)
\label{wardid}\eeq
at one-loop order.

\begin{figure}
\begin{center}
\begin{picture}(40000,8000)
\small
\drawline\fermion[\E\REG](4000,0)[4000]
\drawline\fermion[\W\REG](4000,0)[4000]
\drawline\photon[\N\REG](4000,0)[4]
\drawline\fermion[\E\REG](\photonbackx,\photonbacky)[4000]
\drawline\fermion[\W\REG](\photonbackx,\photonbacky)[1000]
\thicklines
\put(2000,4000){\circle{2000}}
\thinlines
\put(2000,4000){\makebox(0,0){$\Sigma$}}
\drawline\fermion[\W\REG](1000,4000)[1000]
\put(9000,2000){\makebox{$+$}}
\drawline\fermion[\E\REG](15000,0)[4000]
\drawline\fermion[\W\REG](15000,0)[4000]
\drawline\photon[\N\REG](15000,0)[4]
\drawline\fermion[\W\REG](\photonbackx,\photonbacky)[4000]
\drawline\fermion[\E\REG](\photonbackx,\photonbacky)[1000]
\thicklines
\put(17000,4000){\circle{2000}}
\thinlines
\put(17000,4000){\makebox(0,0){$\Sigma$}}
\drawline\fermion[\E\REG](18000,4000)[1000]
\put(20000,2000){\makebox{$+$}}
\drawline\fermion[\E\REG](26000,0)[1000]
\drawline\fermion[\W\REG](26000,0)[4000]
\drawline\photon[\N\REG](26000,0)[4]
\drawline\fermion[\E\REG](\photonbackx,\photonbacky)[4000]
\drawline\fermion[\W\REG](\photonbackx,\photonbacky)[4000]
\thicklines
\put(28000,0){\circle{2000}}
\thinlines
\put(28000,0){\makebox(0,0){$\Sigma$}}
\drawline\fermion[\E\REG](29000,0)[1000]
\put(31000,2000){\makebox{$+$}}
\drawline\fermion[\E\REG](37000,0)[4000]
\drawline\fermion[\W\REG](37000,0)[1000]
\drawline\photon[\N\REG](37000,0)[4]
\drawline\fermion[\E\REG](\photonbackx,\photonbacky)[4000]
\drawline\fermion[\W\REG](\photonbackx,\photonbacky)[4000]
\thicklines
\put(35000,0){\circle{2000}}
\thinlines
\put(35000,0){\makebox(0,0){$\Sigma$}}
\drawline\fermion[\W\REG](34000,0)[1000]
\end{picture}
\end{center}
\begin{center}
\begin{picture}(40000,7000)
\small
\drawline\fermion[\E\REG](0,1000)[3000]
\thicklines
\put(4000,1000){\circle{2000}}
\thinlines
\drawline\fermion[\E\REG](5000,1000)[3000]
\put(4000,1000){\makebox(0,0){$\Sigma$}}
\put(0,2000){\makebox(0,0){$p$}}
\put(8000,2000){\makebox(0,0){$p$}}
\put(10000,1000){\makebox(0,0){$=$}}
\drawline\fermion[\E\REG](12000,1000)[7800]
\drawline\photon[\NE\REG](14000,1000)[3]
\drawline\photon[\NW\FLIPPED](17800,1000)[3]
\put(21000,1000){\makebox(0,0){$+$}}
\drawline\fermion[\E\REG](23000,1000)[6000]
\drawline\photon[\NW\FLIPPED](26000,1000)[2]
\drawline\photon[\NE\REG](26000,1000)[2]
\drawline\photon[\SE\FLIPPED](26000,3500)[2]
\drawline\photon[\SW\REG](26000,3500)[2]
\put(31000,1000){\makebox(0,0){$+~~\dots$}}
\end{picture}
\end{center}
\begin{description}
\small
\baselineskip=12pt
\item[Figure 3.4:] Total one-loop meson self-energy contribution to the
scattering amplitude. The self-energy operator is $\Sigma(p)$.
\end{description}
\end{figure}

The identity (\ref{wardid}) shows that the total sum of the one-loop vertex
corrections shown in Fig. 3.3 and the meson self-energy corrections in Fig. 3.4
can be reduced simply to the calculation of the ghost self-energy (Fig. 3.5).
In fact, it shows that the longitudinal parts of the contributions from the
3-gluon vertex exchange amplitudes (fourth graph in Fig. 3.3) are determined
entirely by the ghost wavefunction renormalization constant, while the other
longitudinal vertex and self-energy corrections cancel each other out. This is
because, first of all, the meson-meson-gluon-gluon vertex is independent of
momentum and so the one-loop vertex function containing this vertex and the
3-gluon vertex (fifth diagram in Fig. 3.3) is a function only of $q$, and hence
is proportional to $q^\mu$. In the Landau gauge, the amplitudes involving this
vertex function therefore vanish by transversality of the gluon propagator,
$q^\mu G^{ab}_{\mu\nu}(q)=0$. Secondly, modulo a common group theoretical
factor, the first three vertex functions in Fig. 3.3 and the scalar self-energy
operators are the same as those in the abelian theory ($G=U(1)$) where the
Slavnov-Taylor identity (\ref{slavtaylor}) is replaced by the simpler Ward
identity $Z_\phi=Z_e$ (as then $f^{abc}=0$ in (\ref{gaugeaction}) and so the
ghost field decouples from the gauge field). This means that these vertex and
self-energy corrections cancel each other. At one-loop order therefore, the
3-gluon vertex renormalization constant $Z_{e,3}^{(1)}$ is determined entirely
from the ghost wavefunction renormalization constant as
\beq
Z_{e,3}^{(1)}=-\tilde Z^{(1)}
\eeq

\begin{figure}
\begin{center}
\begin{picture}(40000,8000)
\small
\drawline\fermion[\W\REG](2000,3500)[2000]
\drawline\fermion[\E\REG](5000,3500)[2000]
\drawline\photon[\N\REG](3500,0)[2]
\put(4000,0){\makebox(0,0){$q$}}
\put(0,4000){\makebox(0,0){$p$}}
\put(7000,4000){\makebox(0,0){$p-q$}}
\thicklines
\put(3500,3500){\circle{3000}}
\thinlines
\put(3500,3500){\makebox(0,0){$\Gamma^{(1)}_{\rm long}$}}
\put(10000,2500){\makebox(0,0){$+$}}
\put(13000,3000){\makebox(0,0){$p-q$}}
\drawline\fermion[\E\REG](13000,2500)[2000]
\drawline\fermion[\E\REG](18000,2500)[2000]
\thicklines
\put(16500,2500){\circle{3000}}
\thinlines
\put(16500,2500){\makebox(0,0){$\Sigma^{(1)}$}}
\put(20000,3000){\makebox(0,0){$p-q$}}
\put(23000,2500){\makebox(0,0){$-$}}
\put(25000,3000){\makebox(0,0){$p$}}
\drawline\fermion[\E\REG](25000,2500)[2000]
\drawline\fermion[\E\REG](30000,2500)[2000]
\thicklines
\put(28500,2500){\circle{3000}}
\thinlines
\put(28500,2500){\makebox(0,0){$\Sigma^{(1)}$}}
\put(32000,3000){\makebox(0,0){$p$}}
\end{picture}
\end{center}
\begin{center}
\begin{picture}(20000,1000)
\put(0,0){\makebox(0,0){$=$}}
\put(3000,500){\makebox(0,0){$p$}}
\multiput(3000,0)(400,0){6}{\line(300,0){100}}
\thicklines
\put(6500,0){\circle{3000}}
\thinlines
\put(6500,0){\makebox(0,0){$\tilde\Pi^{(1)}$}}
\multiput(8000,0)(400,0){6}{\line(300,0){100}}
\put(10000,500){\makebox(0,0){$p$}}
\put(13000,0){\makebox(0,0){$-$}}
\put(16000,500){\makebox(0,0){$p-q$}}
\multiput(16000,0)(400,0){6}{\line(300,0){100}}
\thicklines
\put(19500,0){\circle{3000}}
\thinlines
\put(19500,0){\makebox(0,0){$\tilde\Pi^{(1)}$}}
\multiput(21000,0)(400,0){6}{\line(300,0){100}}
\put(23000,500){\makebox(0,0){$p-q$}}
\end{picture}
\end{center}
\begin{center}
\begin{picture}(30000,7000)
\small
\multiput(0,1000)(400,0){6}{\line(300,0){100}}
\put(0,1500){\makebox(0,0){$p$}}
\thicklines
\put(3500,1000){\circle{2000}}
\thinlines
\put(3500,1000){\makebox(0,0){$\tilde\Pi$}}
\multiput(5000,1000)(400,0){6}{\line(300,0){100}}
\put(7000,1500){\makebox(0,0){$p$}}
\put(10000,1000){\makebox(0,0){$=$}}
\put(13000,1500){\makebox(0,0){$p$}}
\multiput(13000,1000)(400,0){21}{\line(300,0){100}}
\drawline\photon[\NE\REG](15100,1000)[3]
\drawline\photon[\NW\FLIPPED](18900,1000)[3]
\put(21000,1500){\makebox(0,0){$p$}}
\put(24000,1000){\makebox(0,0){$+~~\dots$}}
\end{picture}
\end{center}
\begin{description}
\small
\baselineskip=12pt
\item[Figure 3.5:] The Slavnov-Taylor identity expressing the universality of
the gauge field coupling to the meson and ghost fields at one-loop order.
$\tilde\Pi(p)$ is the ghost self-energy operator.
\end{description}
\end{figure}

The ghost self-energy is given by
\beq\new{\begin{array}{lll}
\tilde\Pi^{ab}(p)&=&\int\frac{d^3k}{(2\pi)^3}~\tilde\Gamma_\mu^{ace}(-p,p-k;k)
\tilde\Gamma_\nu^{dbf}(k-p,p;-k)\tilde
G^{cd}(p-k)G_{ef}^{\mu\nu}(k)\\&=&\frac{4\pi
i}{k}\delta^{ab}p^2C_2(G)M\int\frac{d^3k}{(2\pi)^3}~\frac{(k\cdot p)^2-k^2p^2}
{k^2(k+p)^2(k^2-M^2)}\end{array}}
\eeq
Again the Feynman integrations can be carried out using standard methods
\cite{prao,cswu,gmrr}, and in the anyon limit we get
\beq
\lim_{p\to0}\tilde\Pi^{ab}(p)=\frac{2i}{3k}p^2\delta^{ab}C_2(G)~{\rm sgn}(k)
\eeq
which leads to
\beq
Z_{e,3}^{(1)}=\frac{2}{3k}C_2(G)~{\rm sgn}(k)
\eeq
The contribution to the imaginary part of the amplitude (\ref{vertex}) from the
self-energy and longitudinal vertex corrections in the anyon limit is therefore
given by
\beq
\lim_{q\to0}~{\rm Im}~{\cal V}_{\rm long}(p_1,p_2;q)=\frac{32\pi
}{k}T_R(G)\dim(G)Z_e^{(1)}\frac{\epsilon_{\mu\nu\lambda}p_1^\mu p_2^\nu
q^\lambda}{q^2}
\label{V0}\eeq
which leads to the conformal weight
\beq
\Delta^{(1)}_{\rm long}=\frac{4}{3k^2}T_R(G)C_2(G)~{\rm sgn}(k)
\label{conflong}\eeq

Finally, consider the ladder diagrams depicted in Fig. 3.6. The corresponding
Feynman integrations for the first, fourth and fifth diagrams there are
\beq\new{\begin{array}{l}
A_\Box=i\int\frac{d^3k}{(2\pi)^3}~\tr~(2p_1-q-k)^\mu R^a(2p_2+q-k)^\nu
R^b(2p_1-k)^\lambda R^c(2p_2+k)^\rho R^d\\~~~~~~~~~~~~~~~~~~~~~~~~~\times
S(p_1-k)S(p_2+k)G^{ab}_{\nu\mu}(k-q)G^{cd}_{\lambda\rho}(k)\\A_\bigtriangleup
=\int\frac{d^3k}{(2\pi)^3}~\tr~(2p_2+q+k)^\nu R^a(2p_2+k)^\rho
R^bg^{\mu\lambda}\{R^c,R^d\}S(p_2+k)G^{ac}_{\mu\nu}(q-k)G^{bd}_{\lambda\rho}(k)
\\A_\bigcirc=-i\int\frac{d^3k}{(2\pi)^3}~\tr~g^{\mu\lambda}g^{\nu\rho}
\{R^a,R^b\}\{R^c,R^d\}G_{\mu\nu}^{ac}(q-k)G_{\lambda\rho}^{bd}(k)\end{array}}
\label{gaugeladders}\eeq
The second ``box" diagram in Fig. 3.6 is then given by $A_\Box(p_2\to-(p_2+q))$
while the third ``triangle" graph is $A_\bigtriangleup(p_2\to p_1,q\to-q)$. The
structure of the Feynman integrals in (\ref{gaugeladders}) has been examined
extensively in \cite{szkogsem} where it was shown that each of these ladder
amplitudes vanishes in the anyon limit.

The total conformal weight at one-loop order is therefore completely determined
by the one-loop renormalization of the Chern-Simons coefficient $k$ and is
given as the sum of the weights from the pure gauge vacuum polarization loop
and the longitudinal part of the 3-gluon vertex corrections. Thus the induced
spin up to one-loop order is
\beq
\Delta^{(1)}=\Delta^{(0)}+\Delta_{\rm vac}^{(1)}+\Delta_{\rm
long}^{(1)}=\frac{T_R(G)}{k}-\frac{1}{k^2}T_R(G)C_2(G)~{\rm sgn}(k)
\label{deltatotal1loop}\eeq
which coincides with the leading orders in the asymptotic expansion
(\ref{KZexp}) of the Knizhnik-Zamolodchikov formula (\ref{kzweights}). The
above calculation can also be for the most part extended to 2-loop order using
the 2-loop pure gauge self-energies calculated in \cite{gmrr}. However,
although it is conjectured that the ladder amplitudes vanish to all loop-orders
in the anyon limit \cite{szkogsem}, an explicit demonstration of such a
non-renormalization has yet to be established.

Note that the anomalous spin here depends on the sign of the Chern-Simons
coefficient $k$. Transformations which change the orientation of the 3-manifold
$\cal M$ on which the Chern-Simons gauge theory is defined (such as parity or
time-reversal) change the sign of the Chern-Simons term (\ref{csaction}). The
dependence of the perturbative weights on ${\rm sgn}(k)$ ensures that one can
compensate in the effective action a sign reversal of the Chern-Simons term by
a change in sign of $k$ thus rendering the quantum field theory (\ref{tmgt})
invariant under orientation-reversing isometries of $\cal M$. This feature is
necessary to maintain the covariance of both the topologically massive and
topological Chern-Simons quantum gauge theories. It is also consistent with the
orientation-reversing features of the induced WZNW model on $\partial{\cal M}$
which change the sign of the central charge of the corresponding Kac-Moody
algebra. The relative sign differences between the coefficients of several
Chern-Simons theories is important for the corresponding induced current
algebras where a negative level coefficient would lead to a non-unitary model.

\begin{figure}
\begin{center}
\begin{picture}(40000,7000)
\small
\drawline\fermion[\E\REG](0,1000)[6000]
\drawline\fermion[\E\REG](0,5000)[6000]
\drawline\photon[\N\REG](2000,1000)[4]
\drawline\photon[\N\REG](4000,1000)[4]
\put(7000,2500){\makebox(0,0){$+$}}
\drawline\fermion[\E\REG](8000,1000)[6000]
\drawline\photon[\NW\FLIPPED](12000,1000)[6]
\drawline\fermion[\E\REG](8000,4800)[6000]
\drawline\photon[\NE\REG](10000,1000)[1]
\drawline\photon[\NE\REG](11200,2200)[4]
\put(15000,2500){\makebox(0,0){$+$}}
\drawline\fermion[\E\REG](16000,1000)[8000]
\drawline\fermion[\E\REG](16000,4800)[8000]
\drawline\photon[\NE\FLIPPED](20000,1000)[6]
\drawline\photon[\NW\REG](20000,1000)[6]
\put(25000,2500){\makebox(0,0){$+$}}
\drawline\fermion[\E\REG](26000,1000)[8000]
\drawline\fermion[\E\REG](26000,4800)[8000]
\drawline\photon[\SE\REG](30000,4800)[6]
\drawline\photon[\SW\FLIPPED](30000,4800)[6]
\put(35000,2500){\makebox(0,0){$+$}}
\drawline\fermion[\E\REG](36000,1000)[6000]
\drawline\fermion[\E\REG](36000,4800)[6000]
\drawline\photon[\NW\REG](39000,1000)[3]
\drawline\photon[\NE\FLIPPED](39000,1000)[3]
\drawline\photon[\SW\FLIPPED](39000,4800)[3]
\drawline\photon[\SE\REG](39000,4800)[3]
\end{picture}
\end{center}
\begin{description}
\small
\baselineskip=12pt
\item[Figure 3.6:] One-loop ladder (2-gluon exchange) diagram contribution.
\end{description}
\end{figure}

\subsection{Coupling to Spinor Fields and the Supersymmetric WZNW Model}

It is possible to carry out the same calculation as above when instead of
mesons we couple fermions to topologically massive Yang-Mills theory with the
action
\beq
S_f=\int
d^3x~\left\{\bar\psi_Ai\gamma^\mu\left(\partial_\mu\psi_A-iR_{AB}^aA_\mu^a\psi_
B\right)-m_f\bar\psi_A\psi_A\right\}
\eeq
where $\psi_A$ are 2-component Dirac fields and the (2 + 1)-dimensional
gamma-matrices can be represented by Pauli spin matrices,
$\gamma^0=\sigma^3,\gamma^1=i\sigma^1,\gamma^2=i\sigma^2$ \cite{szkogsem}. In a
3-dimensional spacetime, the fermion mass $m_f$ may be positive or negative for
parity-odd fermion fields. Now the fermion-fermion-gluon vertex is
$-i\gamma_\mu R_{AB}^a$, while the free fermion propagator in momentum space is
\beq
S^{(f)}_{AB}(p)=\langle\bar\psi_A(p)\psi_B(-p)\rangle_\psi=
i(p_\mu\gamma^\mu-m_f)^{-1}\delta_{AB}
\eeq

All of the analysis carries through as above for scalars using standard
identities for Dirac spinors and fermion bilinears in (2 + 1)-dimensions
\cite{szkogsem}, the only differences being that there are no Feynman diagrams
involving a fermion-fermion-gluon-gluon vertex, and that parity-odd fermions
renormalize the Chern-Simons coefficient $k$ at one-loop order
\cite{prao,cswu}. The one-fermion loop contribution to the vacuum polarization
tensor is
\beq
\Pi_{\mu\nu}^{(f)ab}(q)=\int\frac{d^3k}{(2\pi)^3}~{\rm Tr}~R^a\gamma_\mu
S^{(f)}(k)R^b\gamma_\nu S^{(f)}(k-q)
\eeq
where the trace is over both spinor and group indices.
It was shown in \cite{kshift} that this contribution in the infrared limit is
\beq
\lim_{q\to0}\Pi_{\mu\nu}^{(f)ab}(q)=-\delta^{ab}\left\{
\frac{iT_R(G)}{12\pi|m_f|}q^2g_{\mu\nu}^\perp(q)+\frac{{\rm
sgn}(m_f)T_R(G)}{4\pi}\epsilon_{\mu\nu\lambda}q^\lambda\right\}
\label{fermpol}\eeq
The extra parity-odd structure arises from contraction of the spinor vertices
because in (2 + 1)-dimensions the gamma-matrices obey
$\gamma_\mu\gamma_\nu=2g_{\mu\nu}+i\epsilon_{\mu\nu\lambda}\gamma^\lambda$
\cite{szkogsem}. This gives the extra contribution $\Pi_o^{(f)}(0)=T_R(G)~{\rm
sgn}(m_f)/k$ to (\ref{pioddlowenergy}), and thus the Knizhnik-Zamolodchikov
formula up to 1-loop order becomes
\beq
\Delta_f^{(1)}=\frac{T_R(G)}{k}-\frac{T_R(G)}{k^2}\left(C_2(G)~{\rm
sgn}(k)+T_R(G)~{\rm sgn}
(m_f)\right)
\label{deltaferm1loop}\eeq

The formula (\ref{deltaferm1loop}) agrees with the leading orders of the
asymptotic expansion of the conformal dimension
\beq
\Delta_f=\frac{T_R(G)}{k+C_2(G)~{\rm sgn}(k)-T_R(G)~{\rm sgn}(m_f)}
\label{kzsusy}\eeq
Now an interesting feature immediately arises. If we couple the topologically
massive gauge theory to Dirac fermions which transform under the adjoint
representation of $G$, so that $T_R(G)=-C_2(G)$, then we can adjust the mass
parameter of the fermions so that the one-loop contribution to the induced spin
vanishes and (\ref{kzsusy}) coincides exactly with the tree-level result. In
fact, we can adjust the fermion mass $m_f$ and rescale the Grassmann fields so
that the spinor-coupled gauge theory coincides with the supersymmetric
topologically massive gauge theory which is defined by the bulk action
\cite{siegel,brooks,sakai}
\beq\new{\begin{array}{lll}
{\cal S}&=&S_{TMGT}+\frac{1}{2e^2}\int
d^3x~\bar\chi^ai\gamma^\mu\left(\partial_\mu\chi^a+if^{abc}A_\mu^b\chi^c
\right)-\frac{k}{8\pi}\int d^3x~\bar\chi^a\chi^a\\& &~~~+\frac{k}{8\pi}\int
d^3x~i\bar\lambda^a\left(\gamma^\mu\partial_\mu\chi^a-\frac{1}{3}f^{abc}
\gamma^\mu\partial_\mu\lambda^bL^c-\frac{2}{3}f^{abc}\epsilon^{\mu\nu\rho}
\gamma_\nu\partial_\mu\lambda^bA_\rho^c\right)\end{array}}
\label{susytmgt}\eeq
where $\chi^a$ are Majorana fermion fields in the adjoint representation of
$G$, and $L^a$ are auxilliary scalar fields with $\lambda^a$ their (Majorana)
spinor superpartner fields. The Majorana representation of the (2 +
1)-dimensional Dirac algebra can be taken to be $\gamma^0=\sigma^2$,
$\gamma^1=i\sigma^1$, $\gamma^2=i\sigma^3$, and the Majorana spinors
$\chi^a=i\sigma^2(\bar\chi^a)^T$ have 2 real components.

The action (\ref{susytmgt}) is invariant (up to surface terms) under the
infinitesimal $N=1$ supersymmetry transformation
\beq\new{\begin{array}{lll}
\delta A_\mu=-\bar\varepsilon\gamma_\mu\chi+\bar\varepsilon\partial_\mu\lambda
{}~~~~~&,&~~~~~\delta\chi=\epsilon^{\mu\nu\rho}\partial_\nu
A_\rho\gamma_\mu\varepsilon-\gamma^\nu\partial_\nu(L+\gamma^\mu
A_\mu)\varepsilon\\\delta
L=-\bar\varepsilon\chi+\bar\varepsilon\gamma^\mu\partial_\mu\lambda
{}~~~~~&,&~~~~~
\delta\lambda=A\varepsilon+\gamma^\mu A_\mu\varepsilon\end{array}}
\label{susytransf}\eeq
where $\varepsilon$ is a global, complex-valued Grassmann parameter. The
calculation of the one-fermion loop contribution for the Majorana spinor fields
$\chi^a$ is then identical to that above, because in (2 + 1)-dimensions the
Majorana propagator is identical to the Dirac one and thus the Dirac and
Majorana fermion loop contributions are the same. The exact (tree-level)
conformal dimension $\Delta_{SUSY}=C_2(G)/k$ coincides with that of the $N=1$
supersymmetric WZNW model \cite{divec} which is induced by (\ref{susytmgt}) on
the boundary $\partial{\cal M}$ \cite{sakai}. Its exactness at tree-level is a
consequence of the standard non-renormalizations in supersymmetric field
theories where the supersymmetry leads to mutual cancellations between the
bosonic and fermionic loops in perturbation theory. Notice that for the abelian
theory where $G=U(1)$, we have $C_2(G)=0$ and the exactness of the scalar
conformal weight $\Delta_{U(1)}=1/k$ at tree-level follows from the well-known
non-renormalization properties of abelian Chern-Simons gauge theory
\cite{sswu,kshift}. For the supersymmetric abelian theory, the vanishing of
this weight is a consequence of the fact that in the abelian theory $f^{abc}=0$
in (\ref{susytmgt}) and so all fields are decoupled from the gauge field in
this case.

\section{Effects of Gravitational Dressing: Conformal Weights in Topologically
Massive Gravity}

In this Section we shall assume that the spacetime $\cal M$ has a dynamical
metric $g$ with action (\ref{tmg}) and examine the effects of gravitational
dressing on the anomalous spin values of the last Section. As with the
derivation of Liouville theory from topologically massive gravity
\cite{carlip3,kogan3}, we shall find it convenient to study the model
(\ref{tmg}) in the first order formalism for general
relativity\footnote{\baselineskip=12pt It is also possible to carry out this
analysis when the gravitational Chern-Simons action is expressed in terms of
Christoffel connections rather than spin connections \cite{ashworth}.}
\cite{deserx}. We therefore introduce the dreibein fields $e^a=e_\mu^adx^\mu$
which span the frame bundle of $\cal M$. Here and in the following greek
letters will label the spacetime indices (i.e. the components of the local
basis vectors of the tangent space) and latin letters will denote the basis
index of the local Lorentz group $SO(2,1)$ of the tangent bundle. The dreibein
fields are related to the metric $g$ of $\cal M$ by the orthonormality
condition $g^{\mu\nu}e_\mu^ae_\nu^b=\eta^{ab}={\rm diag}(1,-1,-1)$, or
equivalently by the completeness relation
\beq
\eta_{ab}e_\mu^ae_\nu^b=g_{\mu\nu}
\label{dreicompl}\eeq

The action (\ref{tmg}) for topologically massive gravity in the first order
formalism is
\beq
S_{TMG}=S_E+S_{CS}^{(G)}+S_\lambda
\label{TMGaction}\eeq
where
\beq
S_{CS}^{(G)}=\frac{k'}{8\pi}\int_{\cal M}\left(\omega^a\wedge
d\omega^a+\frac{2}{3}\epsilon^{abc}\omega^a\wedge\omega^b\wedge\omega^c\right)
\label{csgrav}\eeq
is the gravitational Chern-Simons action, and
$\omega^a=\epsilon^{abc}\omega^{bc}$ with
$\omega^{ab}=-\omega^{ba}=\omega_\mu^{ab}dx^\mu$ the spin-connection of the
frame bundle of $\cal M$. The Einstein action is
\beq
S_E=\kappa\int_{\cal M}e^a\wedge R^a
\label{einsteinaction}\eeq
where
\beq
R^a=R^a_{\mu\nu}dx^\mu\wedge dx^\nu=d\omega^a+\epsilon^{abc}\omega^b\wedge
\omega^c
\label{curvom}\eeq
is the curvature of the spin-connection $\omega^a$. We have also included in
(\ref{TMGaction}) the Lagrange multiplier term
\beq
S_\lambda=\int_{\cal M}\lambda^a\wedge\left(de^a+2\epsilon^{abc}\omega^b\wedge
e^c\right)
\label{lagrangeterm}\eeq
where the Lagrange multiplier fields $\lambda^a=\lambda_\mu^adx^\mu$ enforce
the constraint which ensures that the usual covariant derivative $\nabla$
constructed from the spin-connection is compatible with the metric $g$ (i.e.
$\nabla e^a=0$) so that the Einstein action in the first order formalism
coincides with the usual one of general relativity.

The ordinary, pure Einstein theory in three-dimensions contains no propagating
degrees of freedom and is a topological field theory. The addition of the
gravitational Chern-Simons term makes the gravitons of the theory massive with
topological mass (\ref{gravmass}). In the infrared limit $\mu\to\infty$
(equivalently $\kappa\to\infty$) this propagating degree of freedom decouples
and the Chern-Simons term in (\ref{TMGaction}) becomes irrelevant (note that
this is opposite to the situation in topologically massive Yang-Mills theory).
Then the kinetic term for the Liouville field in (\ref{liouville}) vanishes and
the gravity theory (\ref{TMGaction}) induces the theory on $\partial{\cal M}$
which describes the moduli space of Riemann surfaces. The calculation of the
conformal weights as in the last Section in this infrared limit will be
associated with those of string theory in the critical dimension when the
induced anomaly vanishes. As before, we shall regulate the gravitational field
theory in the following by computing everything with the coupling to the
topologically massive graviton field $e_\mu^a$, and then take this infrared
limit to recover the observables of the critical string theory. The effective
coupling constant of the topologically massive gravity theory is the
super-renormalizable, dimensionless expansion parameter $\mu/\kappa\sim1/k'$.
The topological graviton mass regulates the infrared divergences of the
Einstein theory and, based on the naive power counting renormalizability of
topologically massive gravity  \cite{desyang,kesz}, we expect all renormalized
quantities to be non-singular.

The action (\ref{TMGaction}) is diffeomorphism invariant (i.e. generally
covariant) and it also possesses a local $SO(2,1)$-invariance defined by
rotations of the dreibein fields. In this context, the model can be viewed as
an $SO(2,1)$ gauge theory for the spin-connection $\omega^a$. Both the dreibein
and Lagrange multiplier fields above transform under the adjoint representation
of this gauge group. With this point of view, we will study the model
(\ref{TMGaction}) perturbatively by expanding the graviton field about a flat
background metric, i.e. we shift the dreibein fields as
\beq
e_\mu^a\to e_\mu^a+\delta_\mu^a~~~~~,
\label{dreishift}\eeq
and view the topologically massive gravity theory as a quantum field theory on
a flat space. Introducing the new variables $\beta^a$ defined by
\beq
\beta^a=\lambda^a+\kappa\omega^a
\label{betadef}\eeq
the topologically massive gravity action becomes
\beq\new{\begin{array}{c}
S_{TMG}=\int d^3x~\left\{\epsilon^{\mu\nu\lambda}\beta_\mu^a\partial_\nu
e_\lambda^a+2(\beta_\mu^\mu\omega_\nu^\nu-\beta_\nu^\mu\omega_\mu^\nu)+
\frac{k'}{8\pi}\epsilon^{\mu\nu\lambda}\omega_\mu^a\partial_\nu\omega_
\lambda^a-\kappa(\omega^\mu_\mu\omega_\nu^\nu-\omega_\nu^\mu\omega_\mu^\nu)
\right.\\\left.+\epsilon^{\mu\nu\lambda}\left(2\epsilon^{abc}\beta_\mu^a
\omega_\nu^be_\lambda^c-\kappa\epsilon^{abc}e_\mu^a\omega_\nu^b\omega_
\lambda^c+\frac{k'}{12\pi}\epsilon^{abc}\omega_\mu^a\omega_\nu^b\omega
_\lambda^c\right)\right\}\end{array}}
\label{TMGshift}\eeq
where here and in the following all repeated indices are understood to be
summed over by contracting with the flat Minkowski metric $\eta^{ab}$. To
obtain the Feynman rules, we note first of all that it is not necessary to fix
the gauge for the spin-connection $\omega^a$, because its quadratic form in
(\ref{TMGshift}) is non-degenerate (due to the topological mass term). Thus we
fix the gauge only for the $e$ and $\beta$ fields, which is done by introducing
a Faddeev-Popov ghost action (\ref{gaugeaction}) for each of them with
$f^{abc}=\epsilon^{abc}$ and $A=e$ or $\beta$ there. The quadratic part of the
action (\ref{TMGshift}) is then
\beq\new{\begin{array}{c}
S_2=\int d^3x~\left\{\epsilon^{\mu\nu\lambda}\beta_\mu^a\partial_\nu
e_\lambda^a+\frac{1}{2\xi_e}(\partial^\mu
e_\mu^a)^2+\frac{1}{2\xi_\beta}(\partial^\mu\beta_\mu^a)^2+\frac{k'}{8\pi}
\epsilon^{\mu\nu\lambda}\omega_\mu^a\partial_\nu\omega_\lambda^a\right.
\\\left.-\kappa
(\omega^\mu_\mu\omega^\nu_\nu-\omega^\mu_\nu\omega^\nu_\mu)+2(\beta_\mu^\mu
\omega_\nu^\nu-\beta_\nu^\mu\omega_\mu^\nu)\right\}\end{array}}
\label{quadgravaction}\eeq
where $\xi_e$ and $\xi_\beta$ are gauge fixing parameters. As before, the
simplest gauge choice is the transverse Landau gauge wherein
$\xi_e=\xi_\beta=0$ and the propagators of the fields are all transverse. In
what follows we shall choose this gauge.

To transform quantities to momentum space, we make the replacements
$\partial_\mu\to ip_\mu$ in (\ref{quadgravaction}). Then the $e\beta$
propagator is
\beq
-i\langle\beta_\mu^i(p)e_\nu^j(-p)\rangle_{e\beta}=-i\eta^{ij}\epsilon_{
\mu\nu\lambda}p^\lambda/p^2
\label{ebetaprop}\eeq
The propagator for the spin connection
\beq
-i\left\langle\omega_\mu^i(p)\omega^j_\nu(-p)\right\rangle_\omega=\frac{4\pi}
{k'}~\Omega_{\mu\nu}^{ij}(p)
\label{omprop}\eeq
is determined as the solution of the equation
\beq
(-i\epsilon_{\mu\rho\lambda}p^\lambda\eta^{ik}-\mu
X_{\mu\rho}^{ik})~\Omega_{\rho\nu}^{kj}(p)=\eta^{ij}\eta_{\mu\nu}
\label{Xieq}\eeq
where
\beq
X_{\lambda\rho}^{kl}=\delta_\lambda^k\delta^l_\rho-\delta^l_\lambda
\delta_\rho^k
\label{Xdef}\eeq
Inverting the equation (\ref{Xieq}), we find after some algebra
\beq\new{\begin{array}{lll}
\Omega_{\mu\nu}^{ij}(p)&=&\frac{\mu(\Lambda_\mu^i(p)\Lambda_\nu^j(p)-\Lambda
_{\mu\nu}(p)\Lambda^{ij}(p)-\Lambda_\mu^j(p)\Lambda_\nu^i(p))}{2(p^2-\mu^2)}
-\frac{(\Lambda_{\mu\nu}(p)\Lambda^{ij}(p)-\Lambda_\mu^j(p)\Lambda
_\nu^i(p))}{2\mu}\\&
&-\frac{1}{2\mu^3}(\Lambda_{\mu\nu}(p)p^ip^j+\Lambda^{ij}(p)p_\mu
p_\nu-\Lambda_\mu^j(p)p^ip_\nu-\Lambda_\nu^i(p)p_\mu p^j)
\\& &-\frac{i(\epsilon_{\mu\nu\lambda}p^\lambda\Lambda^{ij}(p)+
\epsilon^{ijk}p_k\Lambda_{\mu\nu}(p))}{2(p^2-\mu^2)}\\& &-\frac{i}{2\mu^2}
\left(\epsilon^{i\lambda}_{~~\mu}p_\lambda\frac{p_\nu
p^i}{p^2}+\epsilon_\nu^{~\lambda j}p_\lambda\frac{p_\mu
p^i}{p^2}+\epsilon_{\mu\lambda\nu}p^\lambda\frac{p^i
p^j}{p^2}-\epsilon^{i\lambda j}p_\lambda\frac{p_\mu
p_\nu}{p^2}\right)\end{array}}
\label{Xisoln}\eeq
where
\beq
\Lambda_{\mu\nu}(p)=\eta_{\mu\nu}-p_\mu p_\nu/\mu^2
\label{Pidef}\eeq

The $e\omega$ and pure graviton propagators can now be obtained from the
momentum space convolutions
\beq
-i\langle\omega^i_\mu(p)e_\nu^j(-p)\rangle_{e\omega}=2\langle\omega^i_\mu(p)
\omega_\lambda^k(-p)\rangle_\omega
{}~X_{\lambda\rho}^{kl}~\langle\beta_\rho(p)e_\nu^j(-p)\rangle_{e\beta}
\label{omepropconv}\eeq
\beq
D_{\mu\nu}^{ij}(p)\equiv\langle e_\mu^i(p)e_\nu^j(-p)\rangle_e=-4\langle
e_\mu^i(p)\beta_\lambda^k(-p)\rangle_{e\beta}~X_{\lambda\rho}^{kl}~\langle
\omega_\rho^l(p)\omega_\alpha^m(-p)\rangle_\omega
{}~X_{\alpha\sigma}^{mn}~\langle\beta_\sigma^n(p)e_\nu^j(-p)\rangle_{e\beta}
\label{Ddef}\eeq
which, using (\ref{ebetaprop})--(\ref{Pidef}), after some algebra lead to
\beq
\left\langle\omega^i_\mu(p)e_\nu^j(-p)\right\rangle_{e\omega}=-\frac{8\pi
i}{k'\mu}\left(\frac{i\epsilon_{\mu\nu\lambda}p^\lambda}{p^2}\eta^{ij}-\Omega_
{\mu\lambda}^{ij}(p)\delta^{\perp\lambda}_\nu(p)\right)
\label{omeprop}\eeq
\beq\new{\begin{array}{lll}
D_{\mu\nu}^{ij}(p)&=&\frac{16\pi
i}{k'\mu^2}\left(\frac{\epsilon_{\mu\nu\lambda}p^\lambda}{p^2}\eta^{ij}+
\delta_\mu^{\perp\rho}(p)~\Omega_{\rho\lambda}^{ij}(p)\delta_\nu^{\perp\lambda}
(p)+\frac{\mu}{p^2}\left(\delta_\mu^{\perp j}(p)\delta_\nu^{\perp
i}(p)-\delta_\mu^{\perp i}(p)\delta_\nu^{\perp
j}(p)\right)\right)\\&=&\frac{i}{\kappa}\left(\frac{\mu^2}{2p^2(p^2-\mu^2)}
\left\{\left(\frac{p^2}{\mu^2}-2\right)\eta_{\mu\nu}^\perp(p)\eta^{\perp
ij}(p)+\delta_\mu^{\perp i}(p)\delta_\nu^{\perp j}(p)+\delta_\mu^{\perp
j}(p)\delta_\nu^{\perp i}(p)\right\}\right.\\&
&~~~\left.+\frac{i\mu}{4}\frac{p^\lambda}{p^2(p^2-\mu^2)}\left\{\epsilon_{\mu~
\lambda}^{~i}\delta_\nu^{\perp
j}(p)+\epsilon_{\mu~\lambda}^{~j}\delta_\nu^{\perp
i}(p)+\epsilon_{\nu~\lambda}^{~i}\delta_\mu^{\perp
j}(p)+\epsilon_{\nu~\lambda}^{~j}\delta_\mu^{\perp
i}(p)\right\}\right)\end{array}}
\label{gravprop}\eeq
The propagator (\ref{gravprop}) is the usual Deser-Yang graviton propagator
\cite{desyang} obtained from the second order formalism for topologically
massive gravity. The parity-odd structure in (\ref{gravprop}) can lead to a
gravitational analogue of the Aharonov-Bohm effect \cite{gravanyon}. The
Feynman rules for the interaction vertices $\omega^3$, $\omega^2e$ and
$\omega\beta e$ can be read off from the action (\ref{TMGshift}).

In (2 + 1)-dimensions parity-odd matter fields (such as massive fermions or
topologically massive vector bosons) induce gravitational Chern-Simons terms
\cite{vander}. This term describes the central charge of the corresponding
two-dimensional conformal field theory
\cite{Wita,carkog1,wittencentral,kogan2,carlip3}. It was conjectured in
\cite{kogan2} that the gravitational renormalization of $k'$ must coincide with
that predicted by the KPZ theory \cite{pol2}. Here we shall demonstrate the
equivalence between the KPZ (1 + 1)-dimensional results and the perturbative
$1/k'$ expansion in topologically massive gravity at one-loop order by
calculating the one-loop gravitational renormalization of the conformal
dimension of some primary operators in the infrared limit $\kappa\to\infty$
when the topologically massive gravity theory becomes the topological Einstein
one. As discussed in \cite{kogan3}, it is only in the phase with non-zero
vacuum expectation value for the dreibein field that one can treat the quantum
field theory (\ref{TMGshift}) perturbatively. In the topological phase where
$\langle e_\mu^a\rangle=0$, there is no background space-time and the quadratic
approximation required to find the propagators does not exist. This is because
there are only 2 gauge groups, i.e. diffeomorphisms and local $SO(2,1)$ or
$SL(2,\IR)$ rotations, while all 3 of the fields $\beta$, $\omega$ and $e$ must
be gauge-fixed. Thus we expect to get agreement with the KPZ theory only in the
phase with $\langle e_\mu^a\rangle=\delta^a_\mu$. The topological phase may be
related to the breakdown of the KPZ approach for $1<c<25$ strings.

We consider the minimal coupling (i.e. we assume that the spin connection is
torsion free) of the topologically massive gravity theory above to matter. We
are interested here in the gravitationally dressed scalar-coupled topologically
massive gauge theory of the last Section. For simplicity, we consider only the
abelian case. The results below generalize straightforwardly to the non-abelian
case as well. Thus we are interested in coupling to the matter action
\beq\new{\begin{array}{lll}
S_m[A,\phi,e]&=&\int_{\cal
M}d^3x~\left(-\frac{1}{4e^2}\sqrt{g}~g^{\mu\lambda}g^{\nu\rho}F_{\mu\nu}
F_{\lambda\rho}+\frac{k}{8\pi}\epsilon^{\mu\nu\lambda}A_\mu\partial_\nu
A_\lambda\right.\\&
&\left.~~~~~+\sqrt{g}~g^{\mu\nu}[(\partial_\mu-iA_\mu)\phi]^*[(\partial_\nu
-iA_\nu)\phi]-m^2\sqrt{g}~\phi^*\phi\right)\end{array}}
\label{matgravaction}\eeq
where $g=\det[g_{\mu\nu}]$ and $g^{\mu\nu}=\eta^{ab}e^\mu_ae^\nu_b$ with
$e^\mu_a$ the inverse dreibein fields, i.e. $e_a^\mu e^a_\nu=\delta^\mu_\nu$.
The ghost field terms play no role in the following and so will be ignored.
The topologically massive gauge interaction is introduced to induce a non-zero
bare spin $\Delta_0$ for the meson fields and also to study the interplay
between gauge and gravity degrees of freedom. Otherwise, from the KPZ formula,
we expect to obtain a dressed weight $\Delta=0$. We shall see that the
perturbative calculations in this matter-coupled topologically massive gravity
theory are consistent with this. According to the Knizhnik-Zamolodchikov
formula, the bare conformal spin of primary operators in the associated
conformal field theory is
\beq
\Delta_0=1/k
\label{spin0}\eeq

We are interested in the one-loop order gravitational corrections to the
tree-level Aharonov-Bohm amplitude depicted in Fig. 3.1, i.e. we wish to
determine, to order $1/k'$, the transformation of the bare weight (\ref{spin0})
due to the gravitational dressing from the associated Aharonov-Bohm amplitude
as we did before. Upon shifting the dreibein fields as in (\ref{dreishift}),
besides the usual electrodynamical interactions there are infinitely many
orders of the interactions between the graviton and the photon or meson in
(\ref{matgravaction}). This is because the $\sqrt{g}$ term when expanded in
terms of the shifts (\ref{dreishift}) involves infinitely many powers of the
dreibein field $e^a_\mu$. However, to compute the one-loop corrections to the
diagram in Fig. 3.1, we need only consider those gravitational interactions
involving at most 2 graviton fields $e_\mu^a$. Thus it suffices to expand the
metric determinant factors in (\ref{matgravaction}) to quadratic order in the
dreibein fields,
\beq
\sqrt{g}=1-\frac{1}{2}h^\mu_\mu+\frac{1}{4}\left(\frac{1}{2}h^\mu_\mu
h^\nu_\nu-h^\mu_\nu h^\nu_\mu\right)+\dots
\label{sqrtgexp}\eeq
where
\beq
g_{\mu\nu}=\eta_{\mu\nu}+e_\mu^ae_\nu^a-\eta_{\mu a}e_\nu^a-\eta_{\nu
a}e^a_\mu\equiv\eta_{\mu\nu}+h_{\mu\nu}
\label{metricexp}\eeq
is the expansion of the dynamical metric field about the flat background.
Substituting these expansions into the action (\ref{matgravaction}) we find
\beq\new{\begin{array}{lll}
S_m^{(1)}[A,\phi,e]&=&\int
d^3x~\left\{-\frac{1}{4e^2}F^2+|(\partial-iA)\phi|^2-m^2|\phi|^2+
\frac{k}{8\pi}AdA\right.\\&
&~~~-\frac{1}{4e^2}\left(\eta^{\mu\lambda}\eta^{\nu\rho}
e^\alpha_\alpha-2\eta^{\mu\lambda}\eta^{\rho
a}e_a^\nu-2\eta^{\nu\rho}\eta^{\lambda
a}e_a^\mu-\frac{3}{2}\eta^{\mu\lambda}\eta^{\nu\rho}e^\alpha_\sigma
e^\sigma_\alpha+\eta^{\mu\lambda}e^\nu_ae^\rho_a\right.\\&
&~~~+\eta^{\nu\rho}e^\mu_ae^\lambda_a
+4\eta^{\lambda a}\eta^{\rho b}e^\mu_ae^\nu_b-2\eta^{\mu\lambda}\eta^{\rho
a}e^\nu_ae_\alpha^\alpha-2\eta^{\nu\rho}\eta^{\lambda
a}e^\mu_ae_\alpha^\alpha\\&
&~~~\left.+\frac{1}{2}\eta^{\mu\lambda}\eta^{\nu\rho}e^\alpha_\alpha
e^\sigma_\sigma\right)F_{\mu\nu}F_{\lambda\rho}+\left(\eta^{\mu\nu}
e^\lambda_\lambda+2\eta^{\nu a}e_a^\mu-\eta^{\mu\nu}e_\lambda^\rho
e_\rho^\lambda-e^\mu_ae^\nu_a\right.\\& &\left.~~~+2\eta^{\nu a}e_a^\mu
e_\lambda^\lambda+\frac{1}{2}\eta^{\mu\nu}e^\lambda_\lambda
e^\rho_\rho\right)[(\partial_\mu-iA_\mu)\phi]^*[(\partial_\nu-iA_\nu)
\phi]\\& &\left.~~~-\frac{m^2}{4}\left(2e^\mu_\mu+e^\mu_\mu
e^\nu_\nu-3e^\mu_\nu e^\nu_\mu\right)\phi^*\phi\right\}\end{array}}
\label{1loopgravmat}\eeq

{}From (\ref{1loopgravmat}) we can write down the Feynman rules associated with
each of the graviton-matter interactions relevant for the total one-loop
amplitude. The usual electrodynamical interactions and propagators are as in
the last Section. The graviton-photon-photon vertex is
\beq
{\cal F}^{\mu}_{i\nu\rho}(p;q,r)=\frac{i}{2e^2}\left[(r\cdot
p)\delta^\mu_i\eta_{\nu\rho}-\delta^\mu_ir_\nu p_\rho+p^\mu
r_\nu\eta_{i\rho}+2r_ip_\rho\delta^\mu_\nu-2(r\cdot p)\delta^\mu_\nu
\eta_{i\rho}-2p^\mu r_i\eta_{\nu\rho}\right]
\label{gppvertex}\eeq
where $p=q+r$, while the graviton-graviton-photon-photon vertex function is
given by
\beq\new{\begin{array}{lll}
{\cal F}^{\mu\nu}_{ij\lambda\rho}(p,s;q,r)&=&\frac{i}{e^2}\left[2p_ir_j
\delta^\mu_\lambda
\delta^\nu_\rho-2p_ir^\nu\delta^\mu_\lambda\eta_{j\rho}-(p\cdot
r)\delta^\mu_\lambda\delta^\nu_\rho\eta_{ij}+p_\rho
r^\nu\delta^\mu_\lambda\eta_{ij}+p^\mu
r_\lambda\delta^\nu_\rho\eta_{ij}\right.\\& &~~~-p^\mu
r^\nu\eta_{ij}\eta_{\lambda\rho}-\frac{1}{2}(p\cdot
r)\eta_{\lambda\rho}\delta^\mu_i\delta^\nu_j+\frac{1}{2}p_\rho
r_\lambda\delta^\mu_i\delta^\nu_j\\& &~~~\left.+\frac{1}{4}(p\cdot
r)\eta_{\lambda\rho}\eta^{\mu\nu}\eta_{ij}-\frac{1}{4}p_\rho
r_\lambda\eta^{\mu\nu}\eta_{ij}\right]\end{array}}
\label{ggppvertex}\eeq
with $p+s=q+r$. The meson-meson-graviton vertex is
\beq
{\cal E}_i^\mu(p,p';q)=i(\delta^\mu_i[p\cdot p'-m^2/2]+p'^\mu p_i+p^\mu p'_i)
\label{mmgvertex}\eeq
where $q=p-p'$, the meson-meson-graviton-graviton vertex function is
\beq
{\cal E}_{ij}^{\mu\nu}(p,p';q,r)=i\left[\frac{m^2}{4}\left(3\delta_i^\nu
\delta_j^\mu-\delta_i^\mu\delta_j^\nu\right)+\frac{1}{2}p\cdot
p'\left(\delta^\mu_i\delta^\nu_j-2\delta_j^\mu\delta_i^\nu\right)-p'^\mu
p^\nu\eta_{ij}+\delta_j^\nu\left(p_ip'^\mu+p_i'p^\mu\right)\right]
\label{mmggvertex}\eeq
with $q+r=p-p'$, and the meson-meson-photon-graviton vertex is
\beq
{\cal E}_{i\nu}^\mu(p,p';q;r)=-i[\delta^\mu_i(p+p')_\nu+\delta^\mu_\nu(p+p')_i+
\eta_{i\nu}(p+p')^\mu]
\label{mmpgvertex}\eeq
where $q+r=p-p'$. Finally, the meson-meson-photon-graviton-graviton vertex is
given by
\beq
{\cal S}^{\nu\lambda}_{\mu
ij}(p,p';q;r,s)=-i[\delta^\lambda_j\{(p+p')_i\delta^\nu_\mu+\eta_{\mu
i}(p+p')^\nu\}-\delta^\nu_j\delta^\lambda_i(p+p')_\mu-\eta_{ij}(\delta^\nu_\mu
p^\lambda+\delta^\lambda_\mu p'^\nu)]
\label{Svertex2}\eeq
where $q+r+s=p-p'$. For the one-loop gravitational corrections to the
tree-level Aharonov-Bohm amplitude in Fig. 3.1, we can ignore the meson
interactions involving two photon lines.

\subsection{Ward-Takahashi Identities in the Presence of Gravitational
Dressing}

Before carrying through with the calculation of the required Feynman diagrams,
we shall discuss a bit the simplifications and cancellations of certain graphs
which arise from the $U(1)$ gauge invariance of the gravitationally dressed
Maxwell-Chern-Simons theory above. As we saw in Section 2, the standard set of
Slavnov-Taylor identities (or Ward identities in the abelian case) simplified
the calculation of the full one-loop order amplitude by reducing the
calculation of some combinations of diagrams to a single integration (or
cancelling them out in the case $G=U(1)$). The point we wish to make here is
that the usual Ward-Takahashi identities associated with the $U(1)$ symmetry of
quantum electrodynamics through its minimal coupling to a conserved matter
current are still valid when coupled to gravity as above. After integrating the
Maxwell $F^2$ term in (\ref{matgravaction}) by parts, the current to which the
gauge field $A$ is coupled in the gravitationally dressed theory is
\beq
{\cal J}^\mu(x)=\frac{\delta S_m^{({\rm int})}}{\delta A_\mu}=J^\mu(x)+{\cal
G}^\mu(x)
\label{conscurrtot}\eeq
where
\beq
J^\mu=i\sqrt{g}~g^{\mu\nu}[\phi^*(\partial_\nu-iA_\nu)\phi-\phi((\partial_\nu
-iA_\nu)\phi)^*]
\label{conscurrph}\eeq
is the usual meson current, and
\beq
{\cal G}^\mu=\frac{1}{2}\sqrt{g}~g^{\mu\rho}g^{\lambda\nu}{\cal
P}_{\rho\lambda}A_\nu
\label{conscurrgrav}\eeq
is the current from the gravitational interaction. Here
\beq
{\cal P}_{\mu\nu}=g_{\mu\nu}\nabla^2-\nabla_\mu\nabla_\nu
\label{projop}\eeq
is the symmetric, covariant transverse projection operator on the space of
vectors in configuration space and
$\nabla^2=g^{\mu\nu}\partial_\mu(\sqrt{g}\partial_\nu)/\sqrt{g}=g^{\mu\nu}
\nabla_\mu\nabla_\nu$ the scalar Laplacian associated with the spacetime metric
$g$. In the covariant gauge $\eta^{\mu\nu}\partial_\mu A_\nu=0$, the current
conservation law  $\partial_\mu{\cal J}^\mu=0$ follows from taking a derivative
of the usual Euler-Lagrange equations
\beq
0=\frac{\delta S_m}{\delta
A_\mu}=\partial_\nu\left(\sqrt{g}~g^{\mu\lambda}g^{\nu\rho}F_{\lambda\rho}
\right)-\frac{k}{4\pi}\epsilon^{\mu\nu\lambda}F_{\nu\lambda}+J^\mu
\label{ELAeqs}\eeq
for the photon field, and using the Bianchi identity
$\partial_\mu\epsilon^{\mu\nu\lambda}F_{\nu\lambda}=0$. Moreover, it is readily
verified that the gravitational current (\ref{conscurrgrav}) is conserved,
$\partial_\mu{\cal G}^\mu=0$, which follows from symmetry and transversality of
the projection operator (\ref{projop}).

Thus, the meson and gravitational currents which couple to the photon field are
individually conserved. This will lead to a set of Ward-Takahashi identities
for gravitationally corrected Green's functions in the theory. The first
property we wish to establish is the transversality of the gravitationally
corrected photon propagator. Consider the function
\beq
\bar G_{\mu\nu}(p)=G_{\mu\nu}(p)+G_{\mu\lambda}(p)\langle
J^\lambda(p)A_\nu(-p)\rangle+G_{\mu\lambda}(p)\langle{\cal
G}^\lambda(p)A_\nu(-p)\rangle
\label{barGdef}\eeq
where we have Fourier transformed the above operators to momentum space. Here
$G_{\mu\nu}(p)$ is the free photon propagator (\ref{tmgaugeprop}) and the
vacuum correlation functions correspond to time-ordered Green's functions in
the full interacting theory (\ref{matgravaction}). After shifting the graviton
field as in (\ref{dreishift}), the gravitational current decomposes as
\beq
{\cal G}^\mu={\cal G}^{(0)\mu}+{\cal G}^{(h)\mu}
\label{gravcurrdecomp}\eeq
into the sum of a piece ${\cal G}^{(h)\mu}$ associated with the fluctuation
metrics $h_{\mu\nu}$ in (\ref{metricexp}) and a ``flat" part
\beq
{\cal G}^{(0)\mu}=\frac{1}{2}\eta^{\mu\rho}\eta^{\lambda\nu}{\cal
P}_{\rho\lambda}^{(0)}
A_\nu
\label{flatproj}\eeq
where ${\cal
P}_{\mu\nu}^{(0)}=\eta_{\mu\nu}\partial^2-\partial_\mu\partial_\nu$ is the
transverse projection operator on the Minkowski space of vectors. In momentum
space, this projection operator is the usual one ${\cal
P}_{\mu\nu}^{(0)}(p)=p^2\eta_{\mu\nu}^\perp(p)$, so that the flat current
(\ref{flatproj}) is just the transverse part of the gauge field in Minkowski
space. The meson current $J^\mu$ also decomposes into a ``flat'' part
associated with the pure electrodynamical meson-photon interactions and a piece
corresponding to meson-photon-graviton vertices from the $h_{\mu\nu}$ part. By
definition, the full renormalized photon propagator in the interacting quantum
field theory (\ref{matgravaction}) can be written as (see Fig. 4.1)
\beq
G_{\mu\nu}^{\rm ren}(p)\equiv\langle
A_\mu(p)A_\nu(-p)\rangle=G_{\mu\nu}(p)+G_{\mu\lambda}(p)\langle[J^\lambda(p)
+{\cal G}^{(h)\lambda}(p)]A_\nu(-p)\rangle
\label{photonpropren}\eeq
It includes the free photon propagator, the renormalizations from the
scalar-photon vertices, and all the renormalizations from the gravity couplings
in (\ref{matgravaction}). Then substituting the decomposition
(\ref{gravcurrdecomp}) into (\ref{barGdef}) we get
\beq
\bar G_{\mu\nu}(p)=G_{\mu\nu}^{\rm
ren}(p)+p^2G_{\mu\lambda}(p)\eta^{\lambda\rho}\eta^{\alpha\beta}\eta_{\rho
\alpha}^\perp(p)G_{\beta\nu}^{\rm ren}(p)/2
\label{barGren}\eeq
where the second term in (\ref{barGren}) projects out the transverse component
of the complete photon propagator.

\begin{figure}
\begin{center}
\begin{picture}(30000,10000)
\small
\drawline\photon[\N\REG](2000,1000)[3]
\thicklines
\put(2000,5500){\circle{3000}}
\thinlines
\put(2000,5500){\makebox(0,0){$G^{\rm ren}$}}
\drawline\photon[\N\REG](2000,7000)[3]
\put(1000,10000){\makebox(0,0){$p$}}
\put(1000,1000){\makebox(0,0){$p$}}
\put(3000,10000){\makebox(0,0){$\nu$}}
\put(3000,1000){\makebox(0,0){$\mu$}}
\put(6000,5500){\makebox(0,0){$=$}}
\drawline\photon[\N\REG](8000,1000)[9]
\put(10000,5500){\makebox(0,0){$+$}}
\drawline\photon[\N\REG](14000,1000)[4]
\drawline\photon[\N\REG](14000,8000)[2]
\drawline\fermion[\NE\REG](14000,3000)[1500]
\drawline\fermion[\NW\REG](14000,3000)[1500]
\drawline\fermion[\N\REG](15000,4000)[1500]
\drawline\fermion[\N\REG](13000,4000)[1500]
\thicklines
\put(14000,6500){\circle{3000}}
\thinlines
\put(14000,3000){\circle*{500}}
\drawline\gluon[\W\FLIPPED](14000,3000)[1]
\drawline\fermion[\N\REG](12800,5300)[400]
\drawline\gluon[\N\FLIPPED](12800,3000)[2]
\put(15000,2500){\makebox(0,0){$J^\lambda$}}
\put(18000,5500){\makebox(0,0){$+$}}
\drawline\photon[\N\REG](23000,1000)[4]
\drawline\photon[\N\REG](23000,8000)[2]
\thicklines
\put(23000,6500){\circle{3000}}
\thinlines
\put(23000,3000){\circle*{500}}
\drawline\gluon[\W\FLIPPED](23000,3000)[1]
\drawline\fermion[\N\REG](21800,5300)[400]
\drawline\gluon[\N\FLIPPED](21800,3000)[2]
\drawline\gluon[\E\REG](23000,3000)[1]
\drawline\fermion[\N\REG](24200,5300)[400]
\drawline\gluon[\N\REG](24200,3000)[2]
\put(26000,2500){\makebox(0,0){${\cal G}^{(h)\lambda}$}}
\end{picture}
\end{center}
\begin{description}
\small
\baselineskip=12pt
\item[Figure 4.1:] The complete photon propagator $G_{\mu\nu}^{\rm ren}(p)$ in
the gravitationally-dressed topologically massive gauge theory. The spiral
lines denote the dreibein fields $e_\mu^a$, and, as before, wavy lines depict
the photon fields and straight lines the scalar fields. Here the external legs
of the Feynman diagrams are dressed with free propagators $G_{\mu\nu}(p)$, that
on the right-hand side of the equality are contracted into insertions of the
specified currents which are denoted by the solid circles. The first set of
propagators on the right-hand side generated by $J^\lambda$ contain all
initial, bottom-most vertices including both the pure electromagnetic
meson-photon 3- and 4-point interactions as well as their (infinitely-many)
couplings to the graviton field. In the second set only the (infinitely-many)
photon-graviton couplings inserted by ${\cal G}^{(h)\lambda}$ appear in the
bottom-most interaction vertices of the diagrams.
\end{description}
\end{figure}

In a covariant gauge $\eta^{\mu\nu}\partial_\mu A_\nu=0$, the free photon
propagator obeys $p^\mu G_{\mu\nu}(p)=\xi(p^2)p_\nu$ ($\xi(p^2)=0$ in the
Landau gauge). From the canonical commutation relations of the quantum field
theory we have
\beq
[J^0(x),A_\mu(y)]\delta(x^0-y^0)=[{\cal G}^0(x),A_\mu(y)]\delta(x^0-y^0)=0
\label{cancommA0}\eeq
These commutation relations follow from the fact that the canonical momentum
conjugate to the gauge field $A_\mu$ can be taken to be proportional to
$\epsilon^{0\mu\nu}A_\nu$ in the Chern-Simons gauge theory. By current
conservation we have $p_\mu J^\mu(p)=p_\mu{\cal G}^\mu(p)=0$, and so using
(\ref{cancommA0}) and the transversality of the free photon propagator we have
\beq
p^\mu\bar G_{\mu\nu}(p)=p^\mu G_{\mu\nu}(p)=p^\mu G_{\mu\nu}^{\rm ren}(p)
\label{barGlong}\eeq
where the first equality in (\ref{barGlong}) follows from the definition
(\ref{barGdef}) and the second from (\ref{barGren}) using transversality of the
second term there. This means that the longitudinal parts of $G^{\rm
ren}_{\mu\nu}$ and $G_{\mu\nu}$ coincide, and so the longitudinal part of the
full photon propagator is not renormalized. This holds for the corrections from
the photon, meson and graviton fields. In particular, we can equate the various
orders of the perturbative expansions in both coupling constants $1/k$ and
$1/k'$ to deduce that this result holds for each of the pure gravitational,
pure electrodynamical and mixed gravity-electrodynamic corrections, at each
given order of perturbation theory in the gravitational and gauge couplings.
Thus the vacuum polarizations from each set of corrections will be individually
transverse. Denoting the vacuum polarization due to pure gravitational
radiative corrections by $\Pi^{({\rm grav})}_{\mu\nu}(p)$, we therefore have
the transverse decomposition
\beq
\Pi^{({\rm grav})}_{\mu\nu}(p)=\frac{1}{M}\Pi_e^{({\rm
grav})}(p^2)p^2\eta_{\mu\nu}^\perp(p)-
\frac{k}{4\pi}\Pi_o^{({\rm grav})}(p^2)\epsilon_{\mu\nu\lambda}p^\lambda
\label{vacpolgravdecomp}\eeq
into the usual parity even and odd components.

The other gravitationally dressed Ward-Takahashi identity we wish to point out
is the usual relation between the meson-meson-photon vertex function and the
meson self-energy. Consider the full renormalized vertex function
\beq
V_\mu^{\rm ren}(p,q)=\langle A_\mu(q)\phi^*(p-q)\phi(p)\rangle
\label{renvertfn}\eeq
and define
\beq
\Xi^\mu(p,q)=\langle J^\mu(q)\phi^*(p-q)\phi(p)\rangle+\langle{\cal
G}^\mu(q)\phi^*(p-q)\phi(p)\rangle\equiv\Xi^\mu_\phi(p,q)+\Xi^\mu_g(p,q)
\label{calVdef}\eeq
Expanding the metric about a flat background as before and noting that by
definition we have (see Fig. 4.2)
\beq
G_{\mu\nu}(q)\langle[J^\nu(q)+{\cal
G}^{(h)\nu}(q)]\phi^*(p-q)\phi(q)\rangle=V_\mu^{\rm ren}(p,q)
\label{GcalGid}\eeq
it follows that
\beq
G_{\mu\nu}(q)~\Xi^\nu(p,q)=V_\mu^{\rm
ren}(p,q)+q^2G_{\mu\nu}(q)\eta^{\nu\rho}\eta^{\lambda\alpha}\eta_{\rho\lambda}
^\perp(q)V_\alpha^{\rm ren}(p,q)/2
\label{GcalVVren}\eeq
where the second term in (\ref{GcalVVren}) projects out the transverse
component of the vector function $V_\mu^{\rm ren}$.

\begin{figure}
\begin{center}
\begin{picture}(40000,10000)
\small
\drawline\fermion[\E\REG](0,5500)[3000]
\drawline\fermion[\E\REG](6000,5500)[3000]
\drawline\photon[\N\REG](4500,1000)[3]
\thicklines
\put(4500,5500){\circle{3000}}
\thinlines
\put(4500,5500){\makebox(0,0){$V^{\rm ren}$}}
\put(0,6000){\makebox(0,0){$p$}}
\put(9000,6000){\makebox(0,0){$p-q$}}
\put(3500,1000){\makebox(0,0){$q$}}
\put(5500,1000){\makebox(0,0){$\mu$}}
\put(12000,4000){\makebox(0,0){$=$}}
\drawline\fermion[\E\REG](14000,6500)[3000]
\drawline\fermion[\E\REG](20000,6500)[3000]
\drawline\photon[\N\REG](18500,1000)[4]
\drawline\fermion[\NE\REG](18500,3000)[1500]
\drawline\fermion[\NW\REG](18500,3000)[1500]
\drawline\fermion[\N\REG](19500,4000)[1500]
\drawline\fermion[\N\REG](17500,4000)[1500]
\thicklines
\put(18500,6500){\circle{3000}}
\thinlines
\put(18500,3000){\circle*{500}}
\drawline\gluon[\W\FLIPPED](18500,3000)[1]
\drawline\fermion[\N\REG](17300,5300)[400]
\drawline\gluon[\N\FLIPPED](17300,3000)[2]
\put(19500,2500){\makebox(0,0){$J^\lambda$}}
\put(25000,4000){\makebox(0,0){$+$}}
\drawline\fermion[\E\REG](27000,6500)[3000]
\drawline\fermion[\E\REG](33000,6500)[3000]
\drawline\photon[\N\REG](31500,1000)[4]
\thicklines
\put(31500,6500){\circle{3000}}
\thinlines
\put(31500,3000){\circle*{500}}
\drawline\gluon[\W\FLIPPED](31500,3000)[1]
\drawline\fermion[\N\REG](30300,5300)[400]
\drawline\gluon[\N\FLIPPED](30300,3000)[2]
\drawline\gluon[\E\REG](31500,3000)[1]
\drawline\fermion[\N\REG](32800,5300)[400]
\drawline\gluon[\N\REG](32800,3000)[2]
\put(34500,2500){\makebox(0,0){${\cal G}^{(h)\lambda}$}}
\end{picture}
\end{center}
\begin{center}
\begin{picture}(16000,10000)
\small
\put(0,5000){\makebox(0,0){$=$}}
\drawline\fermion[\E\REG](2000,9000)[2000]
\drawline\fermion[\E\REG](7000,9000)[2000]
\drawline\fermion[\E\REG](11000,9000)[2000]
\drawline\fermion[\E\REG](16000,9000)[2000]
\thicklines
\put(5500,9000){\circle{3000}}
\put(5500,9000){\makebox(0,0){$S^{\rm ren}$}}
\put(10000,9000){\circle{2000}}
\put(10000,9000){\makebox(0,0){$\Gamma$}}
\put(14500,9000){\circle{3000}}
\put(14500,9000){\makebox(0,0){$S^{\rm ren}$}}
\put(10000,4500){\circle{3000}}
\put(10000,4500){\makebox(0,0){$G^{\rm ren}$}}
\drawline\photon[\N\REG](10000,1000)[2]
\drawline\photon[\N\REG](10000,6000)[2]
\end{picture}
\end{center}
\begin{description}
\small
\baselineskip=12pt
\item[Figure 4.2:] The complete meson-meson-photon vertex function $V_\mu^{\rm
ren}(p,q)$ in the gravitationally-dressed topologically massive gauge theory.
The first equality depicted is as in Fig. 4.1 above. The second equality is the
usual decomposition of the vertex function into the proper vertex function
$\Gamma_\lambda(p,q)$ dressed with complete propagators.
\end{description}
\end{figure}

{}From the canonical commutation relations of the quantum field theory we have
\beq\new{\begin{array}{c}
[J^0(x),\phi(y)]\delta(x^0-y^0)=-\phi(x)\delta^{(3)}(x-y)~~~,~~~
[J^0(x),\phi^*(y)]\delta(x^0-y^0)=\phi^*(x)\delta^{(3)}(x-y)\cr[{\cal
G}^0(x),\phi(y)]\delta(x^0-y^0)=[{\cal
G}^0(x),\phi^*(y)]\delta(x^0-y^0)=0\end{array}}
\label{cancomm0}\eeq
and along with current conservation these lead to the identities
\beq
q_\mu~\Xi^\mu_\phi(p,q)=S^{\rm ren}(p-q)-S^{\rm
ren}(p)~~~~~,~~~~~q_\mu~\Xi^\mu_g(p,q)=0
\label{calJids}\eeq
where $S^{\rm ren}(p)=\langle\phi^*(p)\phi(-p)\rangle$ is the full renormalized
meson propagator. In a transverse gauge where $q^\mu
G_{\mu\nu}(q)=\xi(q^2)q_\nu$, the contraction of the equation (\ref{GcalVVren})
with $q^\mu$ yields
\beq
q^\mu G_{\mu\nu}(q)~\Xi^\nu(p,q)=q^\mu V_\mu^{\rm ren}(p,q)=\xi(q^2)[S^{\rm
ren}(p-q)-S^{\rm ren}(p)]
\label{qcontrGcalJ}\eeq
where the first equality in (\ref{qcontrGcalJ}) follows from transversality of
the projection operator in the second term in ({\ref{GcalVVren}), while the
second equality follows from (\ref{calJids}).

Now notice that the usual Furry theorem of quantum electrodynamics holds here,
because the quantum field theory (\ref{matgravaction}) is invariant under the
charge conjugation transformation
\beq
(\phi,A_\mu,e_\mu^a)~{\buildrel C\over\longrightarrow}~(\phi^*,-A_\mu,e_\mu^a)
\label{chargeconjtransf}\eeq
This implies that the only non-vanishing Green's functions of the quantum field
theory are those that contain an even number of external photon lines and the
same number of incoming $\phi$ and outgoing $\phi^*$ external lines. It does
not, however, restrict graviton lines. We can therefore define the one-particle
irreducible vertex function $\Gamma_\mu$ by the identity (Fig. 4.2)
\beq
V_\mu^{\rm ren}(p,q)=\eta^{\lambda\nu}G_{\mu\lambda}^{\rm ren}(q)S^{\rm
ren}(p-q)i\Gamma_\nu(p,q)S^{\rm ren}(p)
\label{VGphidef}\eeq
Contracting the equation (\ref{VGphidef}) with $q^\mu$ and using tranversality
of the free photon propagator in (\ref{barGlong}) and the identity
(\ref{qcontrGcalJ}) leads to the usual $U(1)$ Ward-Takahashi identity
\beq
q^\mu\Gamma_\mu(p,q)=\Sigma(p-q)-\Sigma(p)
\label{WTvertexid}\eeq
between the longitudinal part of the irreducible vertex function and the
one-particle irreducible meson self-energy defined by $S^{\rm
ren}(p)^{-1}=S(p)^{-1}+\Sigma(p)$. The relation (\ref{WTvertexid}) holds to all
orders of $1/k$ and $1/k'$ in the electrodynamical and gravitational
perturbative expansions. In particular, it applies to the pure gravitational
corrections as discussed above.

\subsection{Gravitational Contributions to the Vacuum Polarization}

We shall now begin computing the radiative corrections due to gravity to the
Aharonov-Bohm scattering of two charged mesons depicted in Fig. 3.1. We note
first of all that the sum of the amplitudes depicted in Figs. 3.3 and 3.4
vanishes, where the one-loop graviton contributions to the irreducible vertex
function and the meson self-energy are shown in Fig. 4.3. This follows from the
Ward-Takahashi identity (\ref{WTvertexid}) which cancels the longitudinal
contribution of the vertex function with the self-energy contributions (c.f.
Fig. 3.5). The remaining transverse contribution from the vertex function
(which combines with the parity-even part of the photon propagator) is easily
checked to be non-singular in the limit of zero momentum transfer, and so it
does not contribute to the Aharonov-Bohm amplitude. This non-singular behaviour
of renormalized quantities is expected on general grounds because of the
infrared finiteness of the Yang-Mills theory of Section 3 in the Landau gauge
\cite{desyang,kesz}. These simplifications cancel out 22 potential
contributions.

\begin{figure}
\begin{center}
\begin{picture}(50000,7500)
\small
\drawline\fermion[\E\REG](0,5000)[3000]
\thicklines
\put(5000,5000){\circle{4000}}
\thinlines
\put(5000,5000){\makebox(0,0){$\Gamma^{\rm (grav)}$}}
\drawline\fermion[\E\REG](7000,5000)[3000]
\drawline\photon[\N\REG](5000,0)[3]
\put(12000,2500){\makebox(0,0){$=$}}
\drawline\fermion[\E\REG](14000,5000)[10000]
\drawline\gluon[\NE\REG](16100,5000)[2]
\drawline\gluon[\NW\FLIPPED](21900,5000)[2]
\drawline\photon[\N\REG](19000,0)[5]
\put(26000,2500){\makebox(0,0){$+$}}
\drawline\fermion[\E\REG](28000,5000)[10000]
\drawline\photon[\N\REG](33000,0)[5]
\drawloop\gluon[\N 8](30500,7000)
\end{picture}
\end{center}
\begin{center}
\begin{picture}(35000,5000)
\small
\put(0,2500){\makebox(0,0){$+$}}
\drawline\fermion[\E\REG](2000,5000)[10000]
\drawline\photon[\N\REG](7000,0)[5]
\drawline\gluon[\NW\REG](6900,2100)[2]
\put(14000,2500){\makebox(0,0){$+$}}
\drawline\fermion[\E\REG](16000,5000)[10000]
\drawline\photon[\N\REG](21000,0)[5]
\drawline\gluon[\NE\FLIPPED](20900,2100)[2]
\put(28000,2500){\makebox(0,0){$+$}}
\drawline\fermion[\E\REG](30000,5000)[10000]
\drawline\photon[\N\REG](35000,0)[5]
\drawline\gluon[\NE\FLIPPED](35000,1200)[1]
\drawline\gluon[\SE\REG](35000,5000)[1]
\end{picture}
\end{center}
\begin{center}
\begin{picture}(35000,10000)
\small
\put(17500,0){\makebox(0,0){(a)}}
\put(0,4500){\makebox(0,0){$+$}}
\drawline\fermion[\E\REG](2000,7000)[10000]
\drawline\photon[\N\REG](4500,2000)[5]
\drawline\gluon[\NW\FLIPPED](8300,7000)[1]
\drawline\gluon[\NE\REG](4500,7000)[1]
\put(14000,4500){\makebox(0,0){$+$}}
\drawline\fermion[\E\REG](16000,7000)[10000]
\drawline\photon[\N\REG](23500,2000)[5]
\drawline\gluon[\NW\FLIPPED](23500,7000)[1]
\drawline\gluon[\NE\REG](19700,7000)[1]
\put(30000,4500){\makebox(0,0){$+~~\dots$}}
\end{picture}
\end{center}
\begin{center}
\begin{picture}(50000,15000)
\small
\drawline\fermion[\E\REG](0,5000)[3000]
\put(24000,1000){\makebox(0,0){(b)}}
\thicklines
\put(5000,5000){\circle{4000}}
\thinlines
\drawline\fermion[\E\REG](7000,5000)[3000]
\put(5000,5000){\makebox(0,0){$\Sigma^{\rm (grav)}$}}
\put(12000,5000){\makebox(0,0){$=$}}
\drawline\fermion[\E\REG](14000,5000)[10000]
\drawline\gluon[\NE\REG](16100,5000)[2]
\drawline\gluon[\NW\FLIPPED](21900,5000)[2]
\put(26000,5000){\makebox(0,0){$+$}}
\drawline\fermion[\E\REG](28000,5000)[10000]
\drawloop\gluon[\N 8](30500,7000)
\put(41000,5000){\makebox(0,0){$+~~\dots$}}
\end{picture}
\end{center}
\begin{description}
\small
\baselineskip=12pt
\item[Figure 4.3:] Total one-loop gravitational renormalization of (a) the
meson-meson-photon vertex, and (b) the meson propagator.
\end{description}
\end{figure}

We next consider the gravitational corrections to the photon propagator at
one-loop order (Fig. 4.4). The gravitational vacuum polarization tensor is
given by
\beq\new{\begin{array}{c}
\Pi^{({\rm
grav})}_{\mu\nu}(q)=\int\frac{d^3k}{(2\pi)^3}~\left\{\eta^{\lambda\sigma}
\eta^{\rho\kappa}{\cal F}^\alpha_{i\mu\sigma}(k;q,k-q){\cal
F}^\beta_{j\kappa\nu}(-k;q-k,-q)G_{\lambda\rho}(q-k)D_{\alpha\beta}^{ij}(k)
\right.\\\left.+{\cal
F}^{\alpha\beta}_{ij\mu\nu}(k,-k;q,-q)D_{\alpha\beta}^{ij}(k)\right\}
\end{array}}
\label{vacpolgrav}\eeq
This integral is ultraviolet finite and can be evaluated using dimensional
regularization \cite{lerdavan}. As in the Yang-Mills case, the tadpole
contribution (second term in (\ref{vacpolgrav})) doesn't contribute to the
required parity-odd structure. Only the parity odd part of (\ref{vacpolgrav})
contributes to the Aharonov-Bohm part of the amplitude shown in Fig. 3.2, and
it comes from either the $\epsilon$-term in the photon propagator or the
$\epsilon$-terms in the graviton propagator. Contracting (\ref{vacpolgrav})
with $\frac{k}{4\pi}\frac{\epsilon_{\mu\nu\lambda}q^\lambda}{2q^2}$, after some
lengthy algebra this piece is found to be
\beq\new{\begin{array}{lll}
\Pi_o^{({\rm grav})}(q^2)&=&\frac{\pi
e^2}{2k'\mu}\int\frac{d^3k}{(2\pi)^3}~\frac{1}
{k^2(k^2-\mu^2)[(k+q)^2-M^2]}\left\{(2M+\mu/2)q^4-\mu\frac{q^6}{2k^2}\right.\\&
&~~~~~~~~~~+(2M+5\mu/2)(k+q)^4-4Mq^2(k+q)^2\\&
&~~~~~~~~~~+\mu(k+q)^2\left[\frac{(k+q)^4}{2k^2}
-3q^2-\frac{q^2(k+q)^2}{k^2}+\frac{q^4}{k^2}\right]\\&
&~~~~~~~~~~\left.+\frac{1}{2}\mu
k^2\left[q^2-5(k+q)^2\right]-(2M+\mu/2)k^4\right\}\end{array}}
\label{vacpolgravodd}\eeq
The evaluation of this integral has been discussed in \cite{lerdavan} where it
was shown to vanish as $q\to0$ in dimensional regularization. Consequently,
there is no gravitational renormalization of the Chern-Simons gauge coefficient
$k$ and gravitational radiative corrections do not dynamically generate
topological vector boson mass terms at one-loop order. Thus, in contrast with
the pure gauge theory renormalization, the gravitational contributions to the
vacuum polarization do not affect the Aharonov-Bohm interaction, and hence the
gravitationally dressed conformal dimensions, at one-loop order.

\begin{figure}
\begin{center}
\begin{picture}(25000,10000)
\small
\drawline\photon[\N\REG](0,1000)[3]
\thicklines
\put(0,6000){\circle{4000}}
\thinlines
\put(0,6000){\makebox(0,0){$\Pi^{\rm (grav)}$}}
\drawline\photon[\N\REG](0,8000)[3]
\put(5000,6000){\makebox(0,0){$=$}}
\drawline\photon[\N\REG](8000,1000)[10]
\drawline\gluon[\NE\FLIPPED](8000,3000)[2]
\drawline\gluon[\SE\REG](8000,8800)[2]
\put(14000,6000){\makebox(0,0){$+$}}
\drawline\photon[\N\REG](17000,1000)[10]
\drawloop\gluon[\E 8](19100,9000)
\put(26000,6000){\makebox(0,0){$+~~\dots$}}
\end{picture}
\end{center}
\begin{description}
\small
\baselineskip=12pt
\item[Figure 4.4:] Total one-loop gravitational renormalization of the photon
propagator.
\end{description}
\end{figure}

\subsection{Graviton Ladder Exchange Contributions}

We see therefore that the one-loop gravitational renormalization of the
Aharonov-Bohm interaction receives contributions from {\it only} the ladder
diagrams shown in Fig. 4.5. Note that we have depicted only the topologically
inequivalent Feynman diagrams here. There are actually 9 ladder graphs in
total. The ``triangle" diagram in Fig. 4.5 has a counterpart corresponding to
$p_1\leftrightarrow p_2$, $q\to-q$ and the ``box" graph has a partner
corresponding to $p_2\to-(p_2+q)$ (see Fig. 3.6). Also, with the exception of
the last ``circle" diagram, each diagram has a partner associated with
interchanging the photon and graviton lines (equivalently the interchange of
initial and final particles in the diagrams). The situation we observe here is
opposite to that in the case of the topologically massive Yang-Mills theory of
the last Section, where the ladder diagrams vanish in the anyon limit and the
conformal weight renormalization is determined entirely by the gluon loop
self-energy and the gluon renormalization of the ghost propagator. Thus the
perturbative, infrared structure of topologically massive gravity is in this
sense opposite to that of topologically massive gauge theory.

\begin{figure}
\begin{center}
\begin{picture}(40000,7000)
\small
\drawline\fermion[\E\REG](7000,1000)[7000]
\drawline\fermion[\E\REG](7000,5000)[7000]
\drawline\photon[\N\REG](9000,1000)[4]
\drawline\gluon[\N\REG](12000,1200)[3]
\drawline\fermion[\N\REG](12000,1000)[200]
\drawline\fermion[\N\REG](12000,4500)[500]
\drawline\fermion[\E\REG](16000,1000)[8000]
\drawline\fermion[\E\REG](16000,4800)[8000]
\drawline\photon[\SE\REG](20000,4800)[6]
\drawline\gluon[\SW\FLIPPED](20000,4800)[3]
\drawline\fermion[\E\REG](26000,1000)[6000]
\drawline\fermion[\E\REG](26000,4800)[6000]
\drawline\photon[\NW\REG](29000,1000)[3]
\drawline\gluon[\NE\FLIPPED](29000,1000)[1]
\drawline\photon[\SW\FLIPPED](29000,4800)[3]
\drawline\gluon[\SE\REG](29000,4800)[1]
\end{picture}
\end{center}
\begin{description}
\small
\baselineskip=12pt
\item[Figure 4.5:] The topologically inequivalent one-loop gravitational ladder
(1-photon and 1-graviton exchange) diagrams.
\end{description}
\end{figure}

Let us remark at this stage that the simplification of the amplitude to the
evaluation of only the ladder graphs occurs as well in the non-abelian case.
Although the pure electrodynamical Ward-Takahashi identities become the more
complicated Slavnov-Taylor identities discussed in the previous Section, those
identities associated with the gravitational corrections are still valid
because the gravitational current ${\cal G}^\mu$ is once again given by a
(non-abelian) projection operator. Furthermore, the ghost fields associated
with the Yang-Mills gauge fields do not contribute to the gravitational
corrections because the gauge-fixing and ghost terms in the action are
decoupled from the graviton field $e_\mu^a$. The only difference in the
non-abelian case is that the bare conformal dimension $\Delta_0$ of the
associated primary operators changes.

We are interested in the parity odd parts of the ladder diagrams, i.e. the
integrand terms which can contribute to the $\epsilon_{\mu\nu\lambda}p_1^\mu
p_2^\nu q^\lambda$ structure of the full amplitude relevant for the
gravitational renormalization of the Aharonov-Bohm interaction. This structure
comes from contracting the parity-even part of the photon propagator with the
parity odd part of the graviton propagator, and vice versa. As in the pure
gauge theory case, the parity even parts of the diagrams will represent a
renormalization of the Coulomb charge interaction between the charged mesons
(which vanishes in the anyon limit) and of the Pauli magnetic moment
interaction (which is finite in the anyon limit) \cite{kogsem}. To simplify the
integrands of the corresponding Feynman integrals, we use the on-shell
conditions for the initial and final scalar particles, i.e.
$p_1^2=(p_1-q)^2=m^2$ and $p_2^2=(p_2+q)^2=m^2$. We can also exploit the
tranversality in the Landau gauge of the free photon propagator $G_{\mu\nu}(p)$
in both of its indices, and of the free graviton propagator
$D_{\mu\nu}^{ij}(p)$ in all four of its indices. After some very lengthy and
tedious algebra, the Feynman integrals for the parity-odd parts of the ladder
amplitudes depicted in Fig. 4.5 can be simplified to the following expressions,

\beq\new{\begin{array}{lll}
L_\Box&=&-i\int\frac{d^3k}{(2\pi)^3}~(2p_1-q-k)^\mu(2p_2+q-k)^\nu{\cal
E}^\lambda_i(k+p_1-q,p_1-q){\cal E}^\rho_j(p_2+q-k,p_2+q)\\&
&~~~~~~~~~~~~~~~\times
G_{\mu\nu}(q-k)D_{\lambda\rho}^{ij}(k)S(k+p_1-q)S(p_2+q-k)\\&=&-\frac{16e^2M}
{\kappa}\int\frac{d^3k}{(2\pi)^3}~\frac{\epsilon_{\mu\nu\lambda}p_1^\mu
p_2^\nu(q-k)^\lambda}{k^2(k^2-\mu^2)(k-q)^2[(k-q)^2-M^2][k^2+2k\cdot(p_1-q)]}
\\& &~~~~~~~~~~\times\frac{1}{k^2-2k\cdot(p_2+q)}\left\{(k^2+\mu^2)\left[
\frac{3m^2}{2}\left(\frac{5m^2}{6}+k\cdot(p_1-p_2-2q)\right)\right.\right.\\&
&~~~~~~~~~~\left.+\frac{[k\cdot(p_2+q)]^2}{k^2}\left(k\cdot(q-p_1)
-\frac{m^2}{2}
\right)+\frac{[k\cdot(p_1-q)]^2}{k^2}\left(k\cdot(p_2+q)-\frac{m^2}{2}\right)
\right]\\& &~~~~~~~~~~+2\mu^2\left(m^2-\frac{[k\cdot(p_1-q)]^2}{k^2}\right)
\left(m^2-\frac{[k\cdot(p_2+q)]^2}{k^2}\right)\\&
&~~~~~~~~~~\left.+2(k^2-\mu^2)\left[(p_1-q)\cdot(p_2+q)-\frac{[k\cdot(p_1-q)]
[k\cdot(p_2+q)]}{k^2}\right]^2\right\}\end{array}}
\label{ladder1}\eeq

\beq\new{\begin{array}{lll}
L_\bigtriangleup&=&{\cal
E}^{\mu\lambda}_i(p_1,p_1-q)\int\frac{d^3k}{(2\pi)^3}~(2p_2+q+k)^\nu{\cal
E}^\rho_j(p_2,p_2+k)G_{\mu\nu}(q-k)D_{\lambda\rho}^{ij}(k)S(k+p_2)\\&=&
-\frac{2e^2M}{\kappa}\int\frac{d^3k}{(2\pi)^3}~\frac{1}{k^2(k^2-\mu^2)(k-q)^2
[(k-q)^2-M^2](k^2+2k\cdot p_2)}\\&
&~~~~~~~~~~\times\left\{\epsilon_{\mu\nu\lambda}p_1^\mu p_2^\nu
q^\lambda\left[4m^2(k^2+2m^2)+4(p_2\cdot k)(k^2+\mu^2)-\frac{(p_2\cdot
k)^2}{k^2}\left(2k^2+6\mu^2\right)\right]\right.\\&
&~~~~~~~~~~+\epsilon_{\mu\nu\lambda}p_2^\nu q^\lambda k^\mu\frac{(p_1\cdot
k)}{k^2}\left[2(p_2\cdot
k)(3\mu^2-5k^2)-m^2(k^2-5\mu^2)+\frac{4\mu^2}{k^2}(p_2\cdot k)^2\right]\\&
&~~~~~~~~~~+\epsilon_{\mu\nu\lambda}\left(p_1^\mu p_2^\nu k^\lambda-p_1^\mu
q^\lambda k^\nu\right)\left[2(p_2\cdot k)^2-4m^2(k^2+2\mu^2)\right.\\&
&~~~~~~~~~~\left.\left.-4(p_2\cdot k)(k^2+\mu^2)+\frac{2\mu^2}{k^2}(p_2\cdot
k)^2\right]\right\}\end{array}}
\label{ladder2}\eeq

\beq\new{\begin{array}{lll}
L_\bigcirc&=&i{\cal E}_i^{\mu\nu}(p_1,p_1-q){\cal
E}_j^{\rho\lambda}(p_2,p_2+q)\int\frac{d^3k}{(2\pi)^3}~G_{\nu\lambda}(q-k)
D_{\mu\rho}^{ij}(k)\\&=&\frac{2e^2M}{\kappa}\int\frac{d^3k}{(2\pi)^3}~
\frac{1}{k^2(k^2-\mu^2)(k-q)^2[(k-q)^2-M^2]}\left\{\epsilon_{\mu\nu\lambda}
p_1^\mu p_2^\nu q^\lambda(5k^2-2\mu^2)\right.\\&
&~~~~~~~~~~-3\epsilon_{\mu\nu\lambda}p_1^\mu p_2^\nu
k^\lambda(k^2+4\mu^2)+\frac{2}{k^2}(k^2-\mu^2)\epsilon_{\mu\nu\lambda}
q^\lambda\left[p_1^\mu k^\nu(p_2\cdot k)+k^\mu p_2^\nu(p_1\cdot k)\right]\\&
&~~~~~~~~~~\left.-\frac{18}{k^2}\mu^2\epsilon_{\mu\nu\lambda}q^\lambda\left[
p_1^\mu k^\nu(p_2\cdot k)+k^\mu p_2^\nu(p_1\cdot k)\right]\right\}\end{array}}
\label{ladder3}\eeq

The above expressions for the Aharonov-Bohm amplitudes are exact. One
intriguing feature of them is that the integrands which come from contracting
the even part of the photon propagator with the odd part of the graviton
propagator vanish identically. In fact, we can work out the tree-level
gravitational Aharonov-Bohm interaction amplitude (i.e. the imaginary part of
the Feynman diagram in Fig. 3.1 with the photon line replaced by a graviton
line), and after some algebra we find
\beq
{\cal T}^{({\rm grav})}(p_1,p_2;q)^{\rm odd}=i{\cal E}_i^\mu(p_1,p_1-q){\cal
E}_j^\nu(p_2,p_2+q)D_{\mu\nu}^{ij}(q)^{\rm odd}\equiv0
\label{treegrav}\eeq
where we have included only the parity-odd $\epsilon$-terms of the full
graviton propagator (\ref{gravprop}). As (\ref{treegrav}) vanishes identically
(for all ranges of momenta), there is no gravitational Aharonov-Bohm effect.
The one-loop computations above then show that there is no induced
gravitational Aharonov-Bohm effect from the interactions of the topologically
massive graviton field with the meson and photon fields, i.e. the photon field
does not renormalize the gravitational Chern-Simons term. Thus the only effect
of the gravitational dressing in the infrared regime is to change the induced
spin of the meson field due to its interaction with the Chern-Simons gauge
field. Although this gauge interaction is dressed by gravity, the gravitational
interaction is not dressed by the electromagnetic
field\footnote{\baselineskip=12pt Note that this differs from the gravitational
Aharonov-Bohm effect where spinless, non-dynamical point particles in the
presence of a topologically massive gravity field acquire an induced spin
$k'/32\pi^2$ under adiabatical rotation \cite{gravanyon}. This gravitational
Aharonov-Bohm effect is non-perturbative and comes from the abelian, linearized
approximation to the action.}.

As the initial spin of the charged particles here is zero, this result is
consistent with the KPZ formula for the branch which has $\Delta(0)=0$. One can
check that this feature is also true of the higher-loop amplitudes involving
only graviton lines, which is a result of the index contractions which occur in
the integrands of the Feynman integrals. This therefore agrees with the
expansion (\ref{KPZexp}) of the KPZ formula in which each term is at least of
order $1/k$, i.e. there are no terms at any given order of the $1/k'$ expansion
which aren't accompanied by factors of $1/k$. It is only for particles with
non-zero bare spin, such as fermions or vector bosons, that one would find
non-vanishing pure gravitational corrections. In these cases, the non-zero
diagrams would presumably come from the coupling of the spinning fields to the
spin-connection $\omega_\mu^a$ itself.

In the Appendices some of the technical details of the evaluation of the ladder
diagrams (\ref{ladder1})--(\ref{ladder3}) are discussed. Here we simply quote
the final results of their evaluation in the infrared limit. We write each of
the ladder amplitudes above as
\beq
L(p_1,p_2;q)=\frac{\epsilon_{\mu\nu\lambda}p_1^\mu p_2^\nu q^\lambda}{q^2}
{}~{\cal L}(q^2)
\label{ladderLdef}\eeq
where ${\cal L}(q^2)=\sum_n(q^2)^n~{\cal L}^{(n)}$ are Lorentz-invariant
functions whose coefficients ${\cal L}^{(n)}$ depend only on the bare
parameters of the matter-coupled topologically massive gravity theory and the
kinematical invariants $p_1\cdot p_2$, $p_1\cdot q$ and $p_2\cdot q$. Each of
the separate Feynman integrals in (\ref{ladder1})--(\ref{ladder3}) are found to
converge in the infrared limit $e^2,\kappa\to\infty$. Some of these terms
produce naively divergent contributions when multiplied by the overall
coefficient $e^2M/\kappa\sim M^2/\mu$, but these terms cancel out when the
individual integrals are combined in the full amplitude in the small momentum
limit $q\to0$. This is as hoped for, since the final result for the induced
spin should not depend on any mass scale of the model.

The functions ${\cal L}(q^2)$ above are found to be analytic around $q^2=0$,
thus confirming previous arguments about the infrared finiteness of
topologically massive gravity in the Landau gauge \cite{desyang,kesz}. A
non-zero ${\cal L}(q^2=0)$ will produce a singular pole term at $q^2=0$ in
(\ref{ladderLdef}) and hence a non-vanishing contribution to the Aharonov-Bohm
part of the amplitude. After evaluating the Feynman integrals in
(\ref{ladder1})--(\ref{ladder3}) using the techniques sketched in the
Appendices, we get the final results for the ladder amplitudes in the infrared
limit,
\beq\new{\begin{array}{l}
\new{\begin{array}{ll}{\cal L}_\Box(0)=\lim_{q^2\to0}&-\frac{8\pi
i}{kk'm^2q^2(p_1-q)\cdot q(p_2\cdot q)}\left[8q^6(p_1\cdot
p_2)+2q^4\left(m^2(p_2-p_1)\cdot q\right.\right.\\&~~~\left.+(p_1\cdot
p_2)(p_2-4p_1)\cdot q\right)+2q^2\left(m^2(p_1\cdot q)^2+m^2(p_2\cdot
q)^2\right.\\&~~~\left.-3(p_1\cdot p_2)(p_1\cdot q)(p_2\cdot
q)\right)+\pi^2\left(m^2q^2\left\{(p_1\cdot q)(p_2\cdot q)+8(p_2\cdot
q)^2\right.\right.\\&~~~\left.+2(p_1\cdot
q)^2\right\}-m^2q^4p_1\cdot(2q+p_2)+2q^4(p_1\cdot p_2)(3p_2-4p_1)\cdot
q\\&~~~-6m^2(p_1\cdot q)(p_2\cdot q)(p_1+p_2)\cdot q+q^2(p_1\cdot p_2)(p_2\cdot
q)(3p_2-7p_1)\cdot q\\&~~~\left.\left.+8q^6(p_1\cdot p_2)-3(p_1\cdot
p_2)(p_2\cdot q)(p_1\cdot q)(p_1+p_2)\cdot q\right)\right]\end{array}}
\\{\cal L}_\bigtriangleup(0)=\lim_{q^2\to0}-\frac{4\pi i}{kk'q^2(p_2\cdot
q)}\left[6q^4-17(p_2\cdot q)q^2+\pi^2\left(16q^4-3(p_2\cdot q)q^2-18(p_2\cdot
q)^2\right)\right]\\{\cal L}_\bigcirc(0)=0\end{array}}
\label{Lcirc0}\eeq

We see here that, in contrast to the ladder diagrams of topologically massive
gauge theory \cite{szkogsem}, there is a finite contribution from the graviton
ladder exchanges in the infrared limit. That the amplitude $L_\bigcirc$ has no
leading order contributions to the Aharonov-Bohm amplitude can be understood
from the fact that the contraction of the meson-meson-photon-graviton vertices
in (\ref{ladder3}) produces the same sort of Feynman integrals that appear in
the gauge theory ladder amplitudes (\ref{gaugeladders}) which individually
vanish in the anyon limit. The other ladder amplitudes in (\ref{ladder1}) and
(\ref{ladder2}) involve higher-derivative gravitational interaction vertices
leading to Feynman integrals of tensorial rank 4 and 5, whereas in the gauge
theory case the maximum tensorial rank of the Feynman integrals involved is 3
\cite{szkogsem}. These higher rank Feynman integrations are ultimately
responsible for the non-vanishing contribution here in the anyon limit (see the
Appendices), and thus the full gravitationally induced spin comes from the
higher-order couplings of the meson and photon fields to the graviton field. It
is interesting that, unlike the scalar-coupled gauge theory case, the external
charged mesons here must be dynamical quantum fields to generate a
renormalization of the spin. This is similar to the effect of massive spinor
fields coupled to topologically massive Yang-Mills theory where the dynamical
spin of the fermions produces a non-zero contribution. Here the dynamical
nature is needed to induce a non-zero bare spin for the gravitationally
interacting particles.

To determine the corresponding conformal weights, we note that the on-shell
conditions for the external meson lines imply that $p_1\cdot q=-p_2\cdot
q=q^2/2$, and also that $p_1\cdot p_2\to m^2$ in the limit $q^2\to0$. Comparing
with Section 2 we can then identify the induced spin associated with each of
the amplitudes above as
\beq
\Delta_\Box=-\frac{1}{2kk'}\left(14+13\pi^2\right)~~~~~,~~~~~
\Delta_\bigtriangleup=
\frac{29+26\pi^2}{4kk'}~~~~~,~~~~~\Delta_\bigcirc=0
\label{ladderspins}\eeq
Finally, there are the contributions from the remaining ``box" and ``triangle"
graphs which can be obtained from the above amplitudes by permutations of the
external particle lines. The triangle diagrams corresponding to the interchange
of the 2 mesons can be obtained from the above amplitudes by interchanging
$p_1\leftrightarrow p_2$ and reflecting $q\to-q$ in them, and the box diagram
corresponding to crossing the photon and graviton lines in that depicted in
Fig. 4.5 is given by $-L_\Box(p_2\to-(p_2+q))$. The remaining graphs correspond
to the interchange of photon and graviton lines and are given in terms of the
amplitudes as $L(p_2\to-(p_2+q),p_1\to-(p_1-q))$, i.e. by the interchange of
initial and final particles with sign factors appropriate to the change in
direction of the lines. These operations lead to the same Aharonov-Bohm
amplitudes as in (\ref{Lcirc0}), and hence to the same weights. Taking into
account the bare conformal weight from the tree-level diagrams, the total
conformal dimension up to one-loop order is thus
\beq
\Delta_{\rm
grav}^{(1)}=\Delta_0+4\Delta_\Box+4\Delta_\bigtriangleup+\Delta_\bigcirc=
1/k+1/kk'
\label{KPZ1loop}\eeq
which agrees with the leading orders of the expansion (\ref{KPZexp}) of the KPZ
formula.

This is the main result of this paper -- the total one-loop radiative
corrections to the Aharonov-Bohm amplitude in matter-coupled topologically
massive gravity coincides in the anyon limit with the leading orders of the
large-$k'$ expansion of the KPZ formula. It represents a non-trivial
correspondence between observables of the two-dimensional conformal field
theory and the those of its topological membrane description. Notice that, in
contrast to the gauge theory calculation, the conformal dimension
(\ref{KPZ1loop}) does not depend on the sign of $k'$, as expected from both the
topological limit of the topologically massive gravity model and the
identification of the gravitational Chern-Simons coefficient with the central
charge of the $SL(2,\IR)$ current algebra of Liouville theory. The (2 +
1)-dimensional Einstein action can have either sign and it is only the overall
sign of the topologically massive gravity action relative to that of the matter
fields that must be fixed to eliminate potential ghost terms. Thus the quantum
field theory defined by (\ref{TMGaction}) does not depend on ${\rm sgn}(k')$ as
it did in the topological gauge theory case.

\section{Conclusions}

In this Paper we have demonstrated how the spectrum of anomalous dimensions in
conformal field theories can be calculated perturbatively from their associated
topological membrane descriptions. In particular, we have derived the leading
orders of the KPZ conformal weight formula from the infrared limit of
topologically massive gravity coupled to charged scalar fields with non-trivial
anomalous spin. The gauge theory induced spins are essentially determined by
the well-known renormalization of the current algebra level of the WZNW model
in Chern-Simons gauge theory, while the gravitationally dressed dimensions are
determined by the higher-derivative couplings of the anomalous spin of the
charged fields to the graviton. The above discussions have demonstrated how a
perturbative analysis of the topologically massive gravity theory reproduces
several consistent features of the KPZ conformal dimension formula, and how the
three-dimensional perturbative description provides several intuitive
descriptions of the gravity theory in two-dimensions. The coupling of the
topologically massive gravity model to the topologically massive gauge theory
has also demonstrated the interplay between gauge and gravity fields in the
topological membrane description (see Subsection 4.2 above). This illustrates
from a perturbative point of view the relations between world-sheet conformal
fields and the geometry of random surfaces in the induced string theory.

It would be interesting to carry out higher-order checks of the KPZ formula to
check the next orders of its expansion. For instance, the sum of the amplitudes
involving only one graviton line but more than two photon lines should vanish
since (\ref{KPZexp}) contains no order $1/k^nk'$ terms with $n>2$.
Heuristically, the vanishing of gravitationally-corrected ladder amplitudes
with large numbers of photon lines is anticipated because then the insertion of
photon propagators ``softens" the effects of the higher-derivative couplings to
the graviton field and the required Feynman integrations begin to resemble
those of the pure gauge theory. However, to explicitly check these other
aspects of this relation requires evaluation of two- and higher-loop integrals,
which seem nearly intractable in light of the large degree of complexity
already involved at one-loop order. Further properties and observables of the
WZNW and Liouville models could be readily verified at one-loop order.

The calculations presented in this Paper are relevant to the topological
membrane description of string theory. It would be interesting to exploit this
description as a starting point for (world-sheet) modifications of string
theory, especially in light of the recent realization that the conventional
understanding of the (target-space) structure of string theory requires a
modification (M-theory). The membrane approach also suggests geometrical
interpretations of observables in the two-dimensional theories. For instance,
the three-dimensional perturbative calculations produce the gravitationally
dressed spin $\Delta$ with the boundary condition $\Delta(\Delta_0=0)=0$. From
the point of view of the two-dimensional Liouville theory, there is no
immediate reason to choose this branch for the solutions of the KPZ scaling
relations. At the calculational level, the topological membrane approach is
appealing because it employs standard techniques of quantum field theory, such
as perturbative renormalization, to study characteristics of the induced string
theory.

\section*{Acknowledgements}

We would like to thank G. Semenoff and A. Vainshtein for helpful discussions.
The work of G.A.-C. was supported in part by funds provided by the European
Community under contract \#ERBCHBGCT940685. The work of R.J.S. was supported in
part by the Natural Sciences and Engineering Research Council of Canada.

\setcounter{section}{0}
\setcounter{subsection}{0}
\addtocounter{section}{1}
\setcounter{equation}{0}
\setcounter{equnum}{0}
\renewcommand{\thesection}{\Alph{section}}

\section*{Appendix A \ \ \ Evaluation of the Gravitational Ladder Diagrams}

In this Appendix we shall briefly outline the strategy used to evaluate the
Feynman integrals (\ref{ladder1})--(\ref{ladder3}). A partial table of results
for the individual Feynman integrals involved can be found in \cite{szkogsem}
and Appendix B to follow. Here we only discuss the methods used for the sake of
completeness. Although the integrations involved in
(\ref{ladder1})--(\ref{ladder3}) can be evaluated using the usual Feynman
parameter techniques, they are rather cumbersome and can be simplified to
simpler algebraic forms via some manipulations. The exact expressions for many
of the individual Feynman integrals involved, and even their infrared limits,
are extremely complicated and not at all informative. We therefore omit the
detailed expressions and simply discuss the simplifying techniques which were
used to arrive at the small momentum results in (\ref{Lcirc0}). These methods
could serve of use for other perturbative calculations in topologically massive
gravity.

As mentioned in Subsection 4.3 above, using the results of \cite{szkogsem} and
carrying out explicit integrations one can see that most of the surviving
contributions in the anyon limit $M,\mu\to\infty$ come from the higher
derivative contributions in (\ref{ladder1}) and (\ref{ladder2}). We begin by
illustrating why this is so. For brevity, we introduce the following shorthand
notations for the propagators appearing in (\ref{ladder1})--(\ref{ladder3}),
\beq
D_g(q,M)=(k-q)^2-M^2~~~~~,~~~~~D_m(p)=k^2+2k\cdot p
\label{brevdens}\eeq
After shifting $p_1\to p_1-q$, $p_2\to-(p_2+q)$ in (\ref{ladder1}), in the
fourth line of the integrand of the second equality there the term
\beq
{\cal B}=\frac{(k\cdot p_1)^2}{k^2}\frac{(k\cdot p_2)^2}{k^2}
\label{calB1def}\eeq
leads to a rank-5 tensor Feynman integral, i.e. one with five loop momentum
factors $k^{\mu_1}\cdots k^{\mu_5}$ in the numerator of (\ref{ladder1}).
However, most of these tensor components appear contracted with the external
meson momenta, and the tensorial rank can be reduced by the trivial identity
\beq
2k\cdot p=D_m(p)-k^2
\label{trivid}\eeq
which at the same time cancels out some of the denominator factors in
(\ref{ladder1}). The cancellation of denominator factors is desired so that a
minimal number of Feynman parameters will be required at the end for the
explicit integrations.

After successive iterations of (\ref{trivid}) one is left with an algebraically
longer set of Feynman integrals to carry out. There are 3 factors of $k^2$ that
can be cancelled out from this process, and then any remaining $k^2$ factors in
the numerator can be cancelled with denominator factors using
\beq
k^2=D_g(0,\mu)+\mu^2
\label{gravtrid}\eeq
It is from here that most of the contributions arise, because the trivial
identity (\ref{gravtrid}) produces a Feynman integral that is multiplied by an
extra overall factor of $\mu^2$. In the anyon limit, this can therefore lead to
potential contributions which would otherwise be suppressed by factors of
$1/\mu^2$. Such tensorial rank reductions can be used for all of the
higher-rank Feynman integrations appearing in the ladder amplitudes. The
resulting algebra is somewhat more lengthy, but this is traded off in the
simplification that now only Feynman integrals up to tensorial rank 3 which
have fewer numbers of denominator factors need be evaluated.

Again some of the Feynman integrations are still rather complicated because of
large numbers of denominator factors which appear in them. Some of them can be
reduced using the partial fraction decompositions
\beq\new{\begin{array}{c}
(k-q)^{-2}D_g(q,M)^{-1}=[D_g(q,M)^{-1}-D_g(q,0)^{-1}]/M^2\\k^{-2}
D_g(0,\mu)^{-1}=[D_g(0,\mu)^{-1}-D_g(0,0)^{-1}]/\mu^2\end{array}}
\label{partfracs}\eeq
This reduces the Feynman integrations to those with only the denominator
factors $D_g(q,M)$ or $D_g(0,\mu)$, and then the remaining part of the
integrations with the factors in (\ref{partfracs}) are obtained by computing
the $M\to0$ or $\mu\to0$ limits of these resulting integrals. The partial
fraction decomposition (\ref{partfracs}) and any shifts in the loop momentum
$k$ are justified because the (2 + 1)-dimensional Feynman integrations involved
are absolutely convergent. After using the decompositions (\ref{partfracs}) it
is possible to identify after some manipulations whether or not the Feynman
integral will contribute in the anyon limit after identifying the overall power
of the photon and graviton masses in the expressions for the amplitudes. This
can be done by using explicit Feynman parametrizations and asymptotic
expansions of the integrands for $M,\mu\to\infty$.

In this way one is left with a (large) succession of Feynman integrals of no
more than rank-3 to evaluate. However, the remaining higher-rank integrations
still potentially involve more than three distinct denominator factors and
therefore require more than two Feynman parameters for their evaluation. They
can be simplified and effectively reduced to scalar rank integrals using the
elegant Brown-Feynman method which was exploited in the case of one-loop
radiative corrections in three-dimensional electrodynamics in \cite{szkogsem}.
Let us briefly sketch the idea behind this method. As an example, consider the
rank-3 tensor Feynman integral
\beq
J_1^{\mu\nu\lambda}=\int\frac{d^3k}{(2\pi)^3}~\frac{k^\mu k^\nu
k^\lambda}{(k^2)^2D_g(0,\mu)(q-k)^2D_g(q,M)D_m(p_2)}
\label{J13}\eeq
which arises in the evaluation of the ladder amplitudes (\ref{ladder1}) and
(\ref{ladder2}). Using (\ref{partfracs}) we can reduce this to the evaluation
of
\beq
J_2^{\mu\nu\lambda}=\int\frac{d^3k}{(2\pi)^3}~\frac{k^\mu k^\nu
k^\lambda}{k^2D_g(0,\mu)(q-k)^2D_g(q,M)D_m(p_2)}
\label{J23}\eeq
so that
$J_1^{\mu\nu\lambda}=[J_2^{\mu\nu\lambda}-J_2^{\mu\nu\lambda}(\mu=0)]/\mu^2$.
The general structure of the integral (\ref{J23}) will be of the form
\beq
J_2^{\mu\nu\lambda}=a^{\mu\nu}p_2^\lambda+b^{\mu\nu}q^\lambda+c^\mu
s^{\nu\lambda}+c^\nu s^{\mu\lambda}
\label{J2decomp}\eeq
where $a^{\mu\nu}$, $b^{\mu\nu}$ are tensor-valued functions and $c^\mu$ a
vector-valued function of $p_2$ and $q$. The symmetric tensor $s^{\mu\nu}$ is
chosen to project out components of vectors transverse to both $p_2$ and $q$,
i.e. $p_{2,\mu}s^{\mu\nu}=q_\mu s^{\mu\nu}=0$, with the normalization
$s_\mu^\mu=1$. Solving these constraints leads to the explicit form
\beq
s^{\mu\nu}=\eta^{\mu\nu}-\frac{1}{m^2q^2-(p_2\cdot q)^2}\left[m^2q^\mu
q^\nu+q^2p_2^\mu p_2^\nu+(p_2\cdot q)\left(q^\mu p_2^\nu+p_2^\mu
q^\nu\right)\right]
\label{stens}\eeq

To determine the as yet unknown functions $a^{\mu\nu}$, $b^{\mu\nu}$ and
$c^\mu$ above, we first contract both sides of the decomposition
(\ref{J2decomp}) with $p_2^\mu$ and $q^\mu$ to get
\beq
2p_{2,\lambda}J_2^{\mu\nu\lambda}=2m^2a^{\mu\nu}+2(p_2\cdot
q)b^{\mu\nu}~~~~,~~~~2q_\lambda J_2^{\mu\nu\lambda}=2(p_2\cdot
q)a^{\mu\nu}+2q^2b^{\mu\nu}
\label{J2contr}\eeq
Inside the integrand of (\ref{J23}), we then use the identities (\ref{trivid})
and
\beq
2q\cdot k=k^2+q^2-(k-q)^2
\label{qtrid}\eeq
to write the left-hand sides of (\ref{J2contr}) as the sum of rank-2 Feynman
integrals which, with the exception of the one multiplied by $q^2$ from
(\ref{qtrid}), have one less denominator factor. This formally determines the
coefficients $a^{\mu\nu}$ and $b^{\mu\nu}$ in (\ref{J2decomp}) in terms of a
set of rank-2 integrations. The vector function $c^\mu$ is then found from the
contraction
\beq
J^{\mu\nu}_{2~~\nu}=p_{2,\nu}a^{\mu\nu}+q_\nu b^{\mu\nu}+e^\mu
\label{J2selfcontr}\eeq
This contraction eliminates the $k^2$ denominator term in the integrand of
$J_2^{\mu\nu\lambda}$ in (\ref{J23}) and produces a vector-valued integral.
Solving the system of algebraic equations (\ref{J2contr}) and
(\ref{J2selfcontr}) then formally determines the rank-3 Feynman integral
(\ref{J23}) in terms of rank-1 and rank-2 Feynman integrals. The rank-2 Feynman
integrals thus generated can then be evaluated in the same way by writing a
decomposition for them analogous to (\ref{J2decomp}) and solving for them in
terms of vector- and scalar-valued Feynman integrals. Finally, the rank-1
integrations can be solved for in terms of a set of scalar-valued integrals,
most of which have fewer denominator factors in their integrands.

After reducing all tensor Feynman integrals to scalar ones in the ladder
amplitudes (\ref{ladder1})--(\ref{ladder3}), the resulting collection of scalar
integrals can be decomposed further using (\ref{partfracs}). All of these
scalar integrations can then be evaluated using the Feynman parametrizations
\beq\new{\begin{array}{lll}
\frac{1}{a^nb}&=&n\int_0^1dx~\frac{x^{n-1}}{[(1-x)b+xa]^{n+1}}\\
\frac{1}{a^nbc}&=&n(n+1)\int_0^1dx~\int_0^xdy~\frac{y^{n-1}}{[ay+b(x-y)+c(1-x)]
^{n+2}}\end{array}}
\label{feynpars}\eeq
and the (2 + 1)-dimensional Feynman integral identity
\beq
\int\frac{d^3k}{(2\pi)^3}~\frac{1}{(k^2+2k\cdot
p+\alpha)^r}=\frac{\Gamma(r-3/2)}{8\pi^{3/2}\Gamma(r)}\frac{1}{(\alpha-p^2)
^{r-3/2}}
\label{feynintid}\eeq
which are valid for $n\geq1$ and $2r>3$. In (\ref{feynpars}) the denominator
factors $a$, $b$ and $c$ will be various combinations of the meson, photon and
graviton propagator denominator factors which appear in
(\ref{ladder1})--(\ref{ladder3}). The above techniques reduce all Feynman
integrations to scalar-valued ones which require no more than three Feynman
parameters for their explicit evaluation. Some of the Feynman parameter
integrations arising from the three denominator parametrization in
(\ref{feynpars}) turn out to be quite complicated and not explicitly computable
because of the required double integration. They can be simplified, however, in
the infrared limit where $M,\mu\to\infty$. After all of these integrations are
carried out, the individual integrals can be expanded about $q^2=0$. Very few
of these expansions will contribute a singular simple pole term at $q^2=0$.
When combined into the expressions for the ladder amplitudes, we arrive at the
leading singular contributions in (\ref{Lcirc0}). For more details about the
above methods and identities, see \cite{szkogsem}.

\section*{Appendix B \ \ \ Feynman Integrals in the Infrared Limit}

In this Appendix we simply list the results of the Feynman integrations which
contribute non-vanishing terms to the amplitudes
(\ref{ladder1})--(\ref{ladder3}) in the infrared limit. The remaining integrals
that appear were either found to vanish in this limit or can be obtained from
those listed below using the methods discussed in Appendix A and by
permutations of the external meson momenta. These were all evaluated using the
above techniques, and some additional results can be found in \cite{szkogsem}.
As discussed before, many of the Feynman integrals that one encounters vanish
in this regime of the quantum field theory. Below we have only listed the
distinct integrals which produce non-vanishing results in this limit, i.e.
those with singular pieces at $q^2=0$. We have also listed the results for the
parameter values $M=\mu$, since, as mentioned before, the conformal dimensions
obtained from these parts of the amplitudes will be independent of all masses
in the theory.
\bd\new{\begin{array}{l}
\int\frac{d^3k}{(2\pi)^3}~\frac{k^\mu}{D_g(0,\mu)D_g(q,0)D_g(q,M)D_m(p_2)}
=-\frac{\pi i}{8M^3q^2}~q^\mu\\ \biggl. ~~~ \biggr.\\ \new{\begin{array}{l}
\int\frac{d^3k}{(2\pi)^3}~\frac{k^\mu
k^\nu}{D_g(0,\mu)D_g(q,0)D_g(q,M)D_m(p_2)}\\=\frac{\pi
i}{16Mq^2}\left\{\eta^{\mu\nu}+\frac{1}{2m^2}\left(\pi^{-2}+2(p_2\cdot
q)/q^2\right)\left(p_2^\mu q^\nu+q^\mu p_2^\nu\right)+\frac{2}{q^2}q^\mu
q^\nu-\frac{1}{m^2}p_2^\mu p_2^\nu\right\}\end{array}}\\ \biggl. ~~~ \biggr.\\
\new{\begin{array}{l}
\int\frac{d^3k}{(2\pi)^3}~\frac{k^\mu
k^\nu}{k^2D_g(q,0)D_g(q,M)D_m(p_2)}\\=\frac{\pi
i}{16Mq^2}\left\{\left(1-2q^2(1+\pi^{-2})/(p_2\cdot
q)\right)\eta^{\mu\nu}-\frac{3}{2}\left(1+\pi^{-2}\right)\left(p_2^\mu
q^\nu+q^\mu p_2^\nu\right)\right.\\~~~~~~~~~~\left.+\frac{2}{p_2\cdot
q}\left(1+\pi^{-2}\right)q^\mu q^\nu+\frac{q^2}{m^2(p_2\cdot
q)}\left(4(1+\pi^{-2})+(p_2\cdot q)/q^2\right)p_2^\mu
p_2^\nu\right\}\end{array}}\\ \biggl. ~~~ \biggr.\\ \new{\begin{array}{l}
\int\frac{d^3k}{(2\pi)^3}~\frac{k^\mu
k^\nu}{k^2D_g(0,\mu)D_g(q,0)D_g(q,M)D_m(p_2)}\\=\frac{\pi i}{8M^3(p_2\cdot
q)}\left\{(1+\pi^{-2})\eta^{\mu\nu}+\frac{p_2\cdot
q}{2m^2q^2}\left(3+4\pi^{-2}+2(p_2\cdot q)/q^2\right)\left(p_2^\mu q^\nu+q^\mu
p_2^\nu\right)\right.\\~~~~~~~~~~\left.-\frac{1}{q^2}\left(1+\pi^{-2}+(p_2\cdot
q)/q^2\right)q^\mu q^\nu-\frac{1}{m^2}\left(3+2\pi^{-2}\right)p_2^\mu
p_2^\nu\right\}\end{array}}\end{array}}
\ed
\bd\new{\begin{array}{l}
J_2^{\mu\nu\lambda}=-\frac{\pi i}{32Mm^2q^2}~\eta^{\mu\nu}p_2^\lambda\\ \biggl.
{}~~~ \biggr.\\ \new{\begin{array}{lll}
J_1^{\mu\nu\lambda}&=&\frac{\pi i}{16M^3q^2}\left\{\frac{1+\pi^{-2}}{p_2\cdot
q}\left(1+\frac{p_2\cdot q}{q^2+p_2\cdot
q}\right)\eta^{\mu\nu}q^\lambda\right.\\& &~~~~~~~~~~
\left.-\frac{(1+\pi^{-2})}{m^2}\left(1+\frac{q^2}
{p_2\cdot q}+\frac{p_2\cdot q}{q^2+p_2\cdot
q}\right)\eta^{\mu\nu}p_2^\lambda\right\}\end{array}}\end{array}}
\ed

\end{document}